\documentclass[aps,pre,twocolumn,showpacs,preprintnumbers,amsmath,amssymb]{revtex4}
\usepackage{graphicx}
\usepackage{dcolumn}
\usepackage{bm}

\begin{document}

\title{Unfolding the procedure of characterizing recorded ultra low frequency, kHZ and MHz electromagetic anomalies 
prior to the L'Aquila earthquake as pre-seismic ones. Part I}

\author{K. Eftaxias}
\author{L. Athanasopoulou}
\author{G. Balasis}
\author{M. Kalimeri}
\author{S. Nikolopoulos}
\author{Y. Contoyiannis}
\author{J. Kopanas}
\author{G. Antonopoulos}
\author{C. Nomicos}

\affiliation{Section of Solid State Physics, Department of Physics, University of Athens, Panepistimiopolis, Zografos, 
15784, Athens, Greece}
\affiliation{Institute for Space Applications and Remote Sensing, National Observatory of Athens, Metaxa and Vas. 
Pavlou, Penteli, 15236, Athens, Greece}
\affiliation{Department of Electronics, Technological Educational Institute of Athens, Ag. Spyridonos, Egaleo, 12210, 
Athens, Greece}







\begin{abstract}
Ultra low frequency, kHz and MHz electromagnetic anomalies were recorded prior to the L'Aquila catastrophic earthquake that occurred on April 6, 2009. The main aims of this contribution are: (i) To suggest a procedure for the designation of detected EM anomalies as seismogenic ones. We do not expect to be possible to provide 
a succinct and solid definition of a pre-seismic EM emission. Instead, we attempt, through a multidisciplinary analysis, to provide elements of a definition. (ii) To link the detected MHz and kHz EM anomalies with equivalent last stages of the L'Aquila earthquake preparation process. (iii) To put forward physically meaningful arguments to support a way of quantifying the time to global failure and the identification of distinguishing features beyond which the evolution towards global failure becomes irreversible. The whole effort is unfolded in two consecutive parts. We clarify we try to specify not only whether or not a single EM anomaly is pre-seismic in itself, but mainly whether a combination of kHz, MHz, and ULF EM anomalies can be characterized as pre-seismic one. In the present first part, we focus on the detected kHz EM anomalies which play a crucial role to the approach of the above mentioned challenges. More precisely, we try to clearly discriminate this anomaly from the noise. For this purpose, we successively analyze the data in terms of various concepts of entropy and information theory: Shannon $n$-block entropy, conditional entropy, entropy of the source, Kolmogorov-Sinai entropy, $T$-entropy, approximate entropy, fractal spectral analysis, R/S analysis and detrended fluctuation analysis. We argue that this analysis reliably recognize the candidate kHz EM precursor from the noise. This hypothesis is supported by the fact that the crucial precursory symptoms included in the candidate precursors are also hidden in other different, in their nature, complex catastrophic events, as it is predicted by the theory of complex systems.
\end{abstract}

\maketitle

\section{Introduction}

A catastrophic earthquake (EQ) of magnitude Mw = 6.3 occurred on
April 6 2009 (01h 32m 41s UTC) in central Italy. This EQ was very shallow. 
The majority of the damage occurred in the city of L'Aquila.

A vital problem in material science and in geophysics is the identification of precursors of 
macroscopic defects or shocks. The EQs are nothing but physical phenomena, namely large scale fractures,
and science should have some predictive power on their future behaviour as for any physical system. 

Earthquakes physicists attempt to link the available observations to the processes occurring in Earth's crust. 
Fracture induced physical fields allow a real-time monitoring of damage evolution in materials during 
mechanical loading. 
When a material is strained, electromagnetic (EM) emissions in a wide frequency spectrum ranging from 
kHz to MHz are produced by opening cracks, which can be considered as the so-called precursors of general 
fracture. These precursors are detectable both at a laboratory and a geological scale 
(Bahat et al., 2005; Eftaxias et al., 2007a; Hayakawa and Fujinawa,1994, Hayakawa, 1999; Hayakawa 
and Molchanov, 2002). 

Since 1994, a station has been installed and operated at a mountainous site of Zante 
island $(37.76^{o}N-20.76^{o}E)$ in the Ionian Sea (western Greece). The aim of this 
station is the detection of EM precursors.
Clear ultra low frequency (ULF), kHz, and MHz EM precursors have been recently detected over periods ranging 
from a few days to a few hours prior to recent catastrophic EQ that occurred in Greece.  
{\it We emphasize that the detected precursors were associated with EQs: (i) that occurred in land 
(or near coast-line); (ii) were strong, i.e., with magnitude 6 or larger;  and (iii) shallow}
(Contoyiannis et al., 2005; Karamanos et al., 2006). Recent results indicate that the recorded EM precursors contain information 
characteristic of an ensuing seismic event (e.g., Eftaxias et al., 2002, 2004, 2006, 2007; 
Kapiris et al., 2004, 2005; Contoyiannis et al., 2005; Contoyiannis and Eftaxias, 2008; 
Kalimeri et al., 2008; Papadimitriou et al., 2008). 

The L'Aquila EQ occured in land, was very shallow and its magnitude was larger than 6. 
MHz, kHz and ULF EM anomalies were observed before this EQ. Notice, 
an important feature, which is observed both at a laboratory and a geological scale, 
is that the MHz radiation appears earlier than the kHz one (Eftaxias et al., 2002 and references therein). 
The detected anomalies followed this temporal scheme: 

 (i) {\it The MHz EM anomalies were detected on 26 March, 2009 and April 2 2009.}

(ii) {\it The kHz EM anomalies were emerged on April 4, 2009.}

(iii) {\it The ULF EM anomaly was continuously recorded from March 29, 2009 up to April 2, 2009.} 

A question naturally arises is whether the recorded anomalies are seismogenic or not. 

{We emphasize that the used experimental setup gives us the possibility  to specify not only whether or not 
a single kHz, MHz, or ULF EM anomaly is pre-seismic in itself, but mainly whether a combination of such kHz, MHz, and ULF 
anomalies can be characterized as pre-seismic.

Some key questions in this field of research are:

 (i) {\it How we can recognize an EM observation as a pre-seismic one. 
We wonder whether necessary and sufficient criteria have been yet established 
that permit to characterize an EM observation as a real EM precursor.} 
 
(ii) {\it How we can link an individual EM precursor with a distinctive stage of EQ preparation.}
 
(iii) {\it How we can identify precursory symptoms in EM observations that witness that the occurrence of the 
prepared EQ is unavoidable.}

Herein, we try to study the possible seismogenic origin of the recorded anomalies prior to the L'Aquila EQ in the frame of the above mentioned open key questions:

(i) {\it A main aim of this contribution is to suggest a procedure for the 
designation of the detected EM anomalies as seismogenic ones. We do not expect to be possible to provide 
a succinct and solid definition of a pre-seismic EM emission. Instead, we attempt, through a multidisciplinary 
analysis, to provide elements of a definition.}
 
(ii) {\it We attempt to link the detected MHz and kHz EM anomalies with the equivalent two last stages
in the L'Aquila EQ preparation process.}

(iii) {\it We attempt to put forward physically
meaningful arguments to support a way of quantifying the time to global failure and the identification of 
distinguishing features beyond which the evolution towards global failure becomes irreversible.}

Notice, recent studies have provided a relevant experience (e.g., Kapiris et al., 2004; Contoyiannis et al., 2005, 2008; 
Papadimitriou et al., 2008; Eftaxias et al., 2006, 2007). This experience encourages this effort and gives us the possibility to 
verify results of this present study with previous ones.   

In the present Part I of this investigation we restrict our study in the designation of the detected kHz EM anomaly prior to the L'Aquila 
EQ as seismogenic one. In Part II we focus on the remaining key open questions. In the following we present our plan. 

Anomaly in a recorded time series is defined as a deviation from the normal behaviour. 
In order to develop a 
quantitative identification of EM precursors, first concepts of entropy and tools of information theory are 
used in order to identify statistical patterns. 
It is expected that, a significant change in the statistical pattern represents a deviation 
from the normal behaviour revealing the presence of an anomaly. 
Symbolic dynamics provides a rigorous way of looking at ``real'' dynamics.  
First, we attempt a symbolic analysis of experimental data in terms of 
Shannon $n$-block entropy, conditional entropy, entropy of the source, Kolmogorov-Sinai entropy, and $T$-entropy.
It is well-known that the Shannon entropy works best in dealing with systems composed of either independent subsystems or interacting via short-range forces, and whose subsystems can access all the available phase space. For systems exhibiting long-range correlations, memory, or fractal properties, Tsallis' entropy becomes the most appropriate
mathematical tool (Tsallis, 1998; 2009). A central property of the
EQ preparation process is the possible occurrence of coherent large-scale collective with a very rich structure, resulting from the repeated nonlinear interactions among collective with a very rich structure, resulting from the repeated nonlinear interactions among its constituents (Sornette, 1999). Consequently, the Tsallis entropy is an appropriate tool to investigate the launch of EM precursor. The Tsallis entropy is also used.

All the techniques based on symbolic dynamics sensitively discriminate the recorder kHz EM anomalies from the EM background: they are characterized by a significantly lower complexity (or higher organization).  

We analyze the data by means of Approximate Entropy ($ApEn$),
which absolutely refers to the raw data, for comparison reasons. This analysis verify the results of symbolic dynamics.    

The fractal spectral analysis offers additional information concerning the 
signal/noise discrimination. Indeed, it:  (i) Shows that the candidate kHz precursors
follow the fractional Brownian motion (fBm)-model; on the contrary, the EM background rather follows the 1/f-noise model. (ii) Implies that the candidate kHz EM precursor has {\it persistent} behaviour. 
The existence of persistency in the candidate precursors is confirmed by   R/S analysis, while, the suggestion that the anomalies follow the fBm-model is verified by Detrended Fluctuation Analysis

The simultaneous abrupt appearance of both high organization (or low complexity) and persistency in the launched kHz anomalies implies 
that the underlying fracto-electromagnetic process is governed by a positive feedback mechanism (Sammis and Sornette, 2002). Such a mechanism is consistent with a candidate precursor, it can be the result of stress transfer from damaged to intact entities ot it can result from the effect 
of damage in lowering the local elastic stiffness (Sammis and Sornette, 2002). Notice, both result {\it in finite-time-singularity}.  Such a singularity, in terms of ``Benioff'' type cumulative EM energy release , is hidden in the emerged kHz anomalies prior to L'Aquila EQ (see Part II). The existence of finite-time-singularity in the anomalies strongly supports the hypothesis of presence of 
a positive feedback mechanism, which expresses a positive circular causality 
that acts as a growth-generating phenomenon and therefore drives unstable patterns (Telesca and Lasaponara, 2006). The appearance of the property of irreversibility in a probable precursor is one of the 
important components of predictive capability (Morgounov, 2001).

Earthquake occurrence is a complex phenomenon in space and time. 
The field of study of complex systems holds that the dynamics of complex systems is founded on universal principles that may used to describe disparate problems ranging from particle physics to the economics of societies. We verify the presence of crucial symptoms that were appeared in the launched EM into other different, however, in their nature, catastrophic events, namely, epileptic seizures, magnetic storms, solar flares.  

The application of the above mentioned procedure sensitively recognizes and discriminates the candidate kHz EM precursors from the EM background. 
Importantly, it extracts universal catastrophic features from the candidate kHz EM precursors. 

{\it However, in the end of this procedure, we report a negative answer to the question whether the extracted symptoms, which have been well 
established and are rooted in a solid physical base, constitute 
not only a necessary but also a sufficient framework in order to characterize the anomalies as pre-seismic ones}.

More effort is needed, which is presented in the following Part II of this contribution.

The paper is organized as follows. Section 2 briefly describes the configuration of the station. It also presents
the candidate ULF, kHz and MHz EM precursors, and discusses a possible their correlation with magnetic storm,
solar flare or atmospheric activity. Section 3 introduces the notion of symbolic dynamics. 
Section 4 provides a brief overview of (Shannon-like) $n$-block entropy, differential or conditional entropy, 
entropy of the source or limit entropy, and Kolmogorov-Sinai entropy, 
as well as the corresponding application of the aforementioned tools to the data. 
Section 5 provides a brief overview of nonextensive Tsallis entropy and the extracted information from the data. Section 6 briefly introduces the notion of $T$-entropy and presents the associated results. Sections 7 presents the notion of $ApEn$ and the corresponding application to the data. In Sections 8, 9 and 10, the fractal spectral analysis R/S analysis and fractal detrended analysis (DFA) are applied to the data. Section 11 discusses the extracted results in terms of theory of complex systems. Finally, Section 12 summarizes and concludes the paper. The conclusions are finalized with the question: which is the next step?

\section{Data presentation}

Since 1994, a station was installed at a mountainous site of Zante island $(37.76^{o}N-20.76^{o}E)$ in 
the Ionian Sea (western Greece) with the following configuration: (i) six loop antennas detecting 
the three components (EW, NS, and vertical) of the variations of the magnetic 
field at 3 kHz and 10 kHz respectively; (ii) two vertical $\lambda/2$ electric dipoles detecting 
the electric field variations at 41 and 54 MHz respectively ,and (iii) two Short Thin Wire Antennas 
(STWA) of 100 m length each, lying on the Earth's surface, detecting ultra low frequency (ULF) ($<~ 1~ Hz$)
anomalies, at EW and NS direction respectively. The 3 kHz, 10 kHz, 41 MHz, and 54 MHz were selected in order to 
minimize the effects of the sources of man-made noise in the mountain area of the Zante island.
All the EM time-series were sampled at 1 Hz.

A sequence of MHz, kHz and ULF EM anomalies were observed before the L'Aquila EQ, as follows:

{\it (i) MHz EM anomalies}. EM anomalies were simultaneously recorded at 41 MHz and 54 MHz by the electric dipoles on March 26, 2009 and April 2, 2009.
 Fig. 1 shows the excerpts of the recorded anomalies by the 41 MHz electric dipole, which show critical behavior.
Physical powerful arguments by means of criticality seem to shed light on their seismogenic origin. 
Growing evidence suggests that rupture in strongly disordered systems can 
be viewed as a type of critical phenomenon (Herrmann and Roux, 1990; Sornette and Andersen, 1998). 
In the Part II of this contribution, we show that the presented in Fig. 1 excerpts of the recorded MHz emission could be described 
in analogy with a thermal continuous (second order) phase transition. 
In this field of research the reproducibility of results is desirable. 
The resulted critical character of the MHz EM anomalies is striking similar 
to that observed in previously detected  MHz EM seismogenic activities (Contoyiannis et al., 2005; 
Contoyiannis and Eftaxias, 2008). The family of asperities distributed along a fault 
sustains the system. We argue that the emerged MHz anomalies could be triggered by 
fractures in the highly disordered system that surrounded the ``backbone of asperities'' 
of the activated L'Aquila fault (Kapiris et al., 2004; Contoyiannis et al., 2005, Contoyiannis and Eftaxias, 2008).   

{\it (ii) kHz EM anomalies}. A sequence of strong multi-peaked EM bursts, with sharp onsets and ends,
were simultaneously recorded by the 3 kHz and 10 kHz loop antennas on April 4, 2009.  
Fig. 2a shows the recorded EM anomalies by the 10 kHz (E-W) loop antenna. These anomalies are launched over a quiescence period concerning the detection of EM disturbances at the kHz frequency band (Fig. 2b).
Fig. 3 depicts magnified images of the excerpts N, B1, B2 and B3 that are shown in fig. 2 (a). 
We argue that their seismogenic origin is strongly supported
by a multidisciplinary analysis. In recent works we suggested that the kHz EM precursors
are rooted in the fracture of asperities of the activated fault
(Kapiris et al., 2004; Contoyiannis et al., 2005;  Eftaxias et al., 2007b,
Eftaxias et al., 2008, Contoyiannis and Eftaxias, 2008;
Papadimitriou et al., 2008). 
The present analysis is consistent with the hypothesis that the detected kHz EM anomalies 
are possibly associated with the fracture of asperities that were distributed along the L'Aquila fault.    
  
{\it (iii) ULF EM anomaly}. An ULF EM anomaly was recorded by the STWA sensors (Fig. 4). 
The daily pattern of the ULF recordings during the normal period, i.e., 
far from the EQ occurrence, follows a rather periodical variation which is characterized by 
the existence of a clear minimum. Someone finds a clear alteration of the normal daily profile
as the shock approaches; the ``depth'' of minimum is significantly decreases. 
The curve attains its normal shape a few days after the event. 
Notice, more clear but similar anomalies were also detected prior to significant EQs in Greece.
They had the same morphology and also emerged a few days before the ensuing EQ 
(Eftaxias et al., 2000, 2002, 2004; Karamanos et al., 2006).  

The appearance of this anomaly may support the hypothesis of a relationship between processes 
produced by increasing tectonic stresses in the Earth's crust and attendant EM interactions between 
the crust and ionosphere. 

We emphasize that we have expressed very clearly our point of view that the
occurrence of an anomaly in the output of a sensor does not qualify
it as a precursor (Karamanos et al., 2006; Papadimitriou et al., 2008). The morphology of the 
recorded ULF signal does not give us the possibility to evaluate their 
seismogenic origin in terms of a rather austere set of criteria as it happens in the 
cases of the kHz and MHz EM activities. The potential seismogenic origin of 
the recorded ULF signal is only based on empirical arguments from 
our side, thus, we consider this anomaly as hypothetical precursory ULF EM emission.

We note that all the recorded EM anomalies have been recorded during a quiet period in terms of
magnetic storm, solar flares and atmospheric activity. The consecutive appearance of ULF, MHz and kHz 
EM anomalies in a time interval a few days prior to 
the L'Aquila EQ  occurrence rather excludes the 
possibility that the recorded anomalies were man-made noises. 

\section{ Fundamentals of symbolic dynamics}
For reasons of completeness and for later use, we compile here the basic points of symbolic dynamics.

Symbolic time series analysis is a 
useful tool for modelling and characterization of nonlinear dynamical systems (Voss et al., 1996).
It provides a rigorous way of looking at "real" dynamics with finite precision
(Hao, 1989, 1991; Kitchens, 1998; Karamanos and Nicolis, 1999). It is a way of coarse-graining or reduction of description.
 
The basic idea is quite simple. One divides the phase space into a finite number of partitions 
and labels each partition with a symbol (a letter from some alphabet). 
Instead of representing the trajectories by infinite sequences of numbers-iterates from a 
discrete map or sampled points along the trajectories of a continuous flow, 
one watches the alteration of symbols. Of course, in so doing, one loses an amount of detailed 
information, but some of the invariant, robust properties of the dynamics may be kept, e.g. periodicity, 
symmetry, or chaotic nature of an orbit (Hao, 1991). 

In the frame of symbolic dynamics the time series are transformed into symbolic sequences by using very 
few symbols based on an appropriate partition. After symbolization, the next step is the construction of 
symbols sequences (words in the symbolic-dynamics literature) from the symbol series by collecting
groups of symbols together in temporal order. 

To be more precise, the simplest possible coarse-graining of a time series is given by choosing a
threshold $C$ (usually the mean value of the data considered) and assigning the symbols
``1'' and ``0'' to the signal, depending on whether it is above or below the threshold (binary partition). 
Thus, we generate a symbolic time series from a 2-letter ($\lambda=2$) alphabet (0,1), e.g. $0110100110010110\ldots$.  
We read this symbolic sequence in terms of distinct consecutive ``blocks'' (words) of length $n=2$. 
In this case one obtains $01/10/10/01/10/01/01/10/\ldots$. We call this reading procedure ``lumping''. 
The number of all possible kinds of words is $\lambda^{n}=2^{2}=4$, namely 00, 01, 10, 11. 

The required probabilities for the estimation of an entropy, $p_{00}$, $p_{01}$, $p_{10}$, $p_{11}$ 
are the fractions of the blocks (words) 00, 01, 10, 11 in the symbolic time series,
namely, 0, 4/16, 4/16, and 0, correspondingly. Based on these probabilities we can estimate, 
for example, the probabilistic entropy measure $H_S$ introduced by Shannon (1948):

\begin{equation}
H_S = - \sum p_i \ln p_i \,
\end{equation}

where $p_{i}$ are the probabilities associated with the microscopic configurations.

Various tools of information theory and entropy concepts are used to identify statistical 
patterns in the symbolic sequences, where the dynamics
of the original (under analysis) system has been projected.
{\it For anomaly detection, it suffices that a detectable change in the pattern represents a 
deviation of the nominal behaviour from an anomalous one (Graben and Kurths, 2003).} Recent literature has reported 
novel methods for anomaly detection in complex dynamical systems, which relies on 
symbolic time series analysis. Entropies depending on the word-frequency distribution 
in symbolic sequences are of special interest, extending Shannon's classical definition of 
the entropy and providing a link between dynamical systems and information theory. 
These entropies take a large (small) value if there are many (few) kinds of patterns, therefore, 
it decreases while the organization of patterns is increasing. In this way, 
these entropies can measure the complexity of a signal.

{\it It is important to note that one cannot find an optimum organization or complexity measure (Kurths et al., 1995). 
We guess that a combination of some such quantities which refer to different aspects, such as structural 
or dynamical properties, seems to be the most promising way. Thus, several well-known techniques have been 
applied to extract EM precursors probably hidden in kHz EM time series.}

We analyze the EM time series by the following techniques which are based on symbolic dynamics:
(i) Shannon $n$-block entropy; (ii) differential or conditional entropy; (iii) limit entropy,
(iv) Kolmogorov-Sinai entropy; (v) $T$-entropy, and finally (vi) Tsallis entropy.  . 

\section{The concept of dynamical (Shannon-like) $n$-block entropies} 

Block entropies, depending on the word-frequency distribution, are of special interest, extending Shannon's classical definition
of the entropy of a single state to the entropy of a succession of states (Nicolis and Gaspard, 1994). Generally, these entropies take a 
large (small) value if there are many (few) kinds of patterns, therefore, it decreases while the organization of 
patterns is increasing. In this way, the block entropy can measure the complexity of a stationary signal.

Symbol sequences, $\{A_{1}...A_{n}...A_{L}\}$, are composed of letters from an alphabet consisting 
of $\lambda$ letters $\{A^{(1)}, A^{(2)}...A^{(\lambda)}\}$.
An English text is written on an alphabet consisting of 26 
letters $\{A, B, C . . . X, Y, Z\}$. 

A word of length $n < L$, $\{A_{1}...A_{n}\}$, is defined by a substring of length $n$ taken
from $\{A_{1}...A_{n}...A_{L}\}$. The total number of different words of length $n$ which exists in is 

$N_{\lambda n} = \lambda ^n$.

We specify that the symbolic sequence is to be read in terms of distinct consecutive ``blocks'' (words) of length $n$,

\begin{equation}
... \underbrace{A_{1}...A_{n}}_{B_{1}} 
\underbrace{A_{n+1}...A_{2 n}}_{B_{2}} ... 
\underbrace{A_{j n+1}...A_{(j+1) n}}_{B_{j+1}}...
\end{equation} 

As it is said, we call this reading procedure {\it lumping}. {\it Gliding} is the reading of the symbolic sequence using a {\it moving 
frame}. 
It has been suggested that in some cases the entropy analysis by lumping is much more sensitive than 
the classical entropy analysis by gliding (Karamanos, 2000, 2001). 

The probability of occurrence of a block $A_{1}...A_{n}$, denoted $p^{(n)}(A_{1},...,A_{n})$, 
is defined by the fraction  

\begin{equation}
\frac{\mbox{No. of blocks,}\quad A_{1}...A_{n}, \quad \mbox{encountered when lumping}}{\mbox{total No. of blocks}}
\end{equation}

starting from the beginning of the sequence. 

The following quantities characterize the information content of the symbolic sequence (Khinchin, 1957; 
Ebeling and Nicolis, 1992):

(i) \textbf{The Shannon $n$-block entropy} 

Following Shannon's approach (Shannon, 1948) the $n$-block entropy, $H_n$, is given by 

\begin{equation}    
H(n) = -\sum_{(A_{1},...,A_{n})}p^{(n)}(A_{1},...,A_{n})\cdot
\ln p^{(n)}(A_{1},...,A_{n})
\end{equation}

{\it The entropy $H_n$ is a measure of uncertainty and gives the average amount of information necessary
to predict a sub-sequence of length $n$.}

(ii) The Shannon $n$-block entropy per letter

This entropy is defined by

\begin{equation}
h^{(n)} = \frac{H(n)}{n}.
\end{equation}

These entropies may be interpreted as the average uncertainty per letter of an $n$-block.

(ii) \textbf{The conditional entropy}

From the Shannon $n$-block entropies we derive the conditional (dynamic) entropies by the definition

\begin{equation}
h_{(n)} = H(n+1) - H(n). 
\end{equation}

{\it The conditional entropy $h_n$ measures the uncertainty of predicting a state one step in the future,
provided a history consisting of $n$-states, namely, the present state and the previous $n-1$ states.}  

Predictability is measured by conditional entropies. For Bernoulli sequences we have maximal uncertainty 

\begin{equation}
h_{(n)} = log(\lambda). 
\end{equation}

Therefore we define the difference

\begin{equation}
r_n=log(\lambda)-h_n
\end{equation}

as the average predictability of the state following after a measured $n$-trajectory. {\it In other words, predictability 
is the information we get by exploration of the next state in the future in comparison to the available knowledge.}
 
We use in most cases $\lambda$ as the units of the logarithms. In these units the maximal 
uncertainty/predictability is one (Ebeling, 1997). 

In general our expectation is that any long-range memory decreases the conditional entropies 
and improves the changes for predictions.

(iii) \textbf{The entropy of the source}

A quantity of particular interest is the entropy of the source defined as 

\begin{equation}
h = \lim_{n \rightarrow \infty} h_{(n)} =  \lim_{n \rightarrow \infty} h^{(n)}
\end{equation}

The limit entropy $h$ is the discrete analog of Kolmogorov-Sinai entropy. 
{\it It is the average amount of information necessary to predict the next symbol when being informed 
about the complete pre-history of the system}. Since positive Kolmogorov-Sinai entropy implies the 
existence of a positive Lyapunov exponent, it is an important measure of chaos.

\subsection{Application to the data}

The upper panel in Fig. 5 shows the entropies performed by {\it lumping}. We recall that {\it lumping}. Is the reading of the symbolic sequence by taking {\it portions}, as opposed to {\it gliding} where one has essentially a {\it moving frame}. The lower panel depicts the entropies estimated by gliding. We observe that both two methods of reading lead to consistent results.

{\it The Shannon $n$-block entropy}. Figs. 5a depict the Shannon $n$-block entropy, $H_n$, as a function of the word length $n$ for the time windows N, B1, B2, and B3 (see Fig. 3). The noise N is characterized by significantly larger 
$H_n$-values. 

{\it This means that the average amount of information necessary
to predict a sub-sequence of length $n$ is larger in the noise than in the bursts B1, B2, and B3 is}. 

{\it The Shannon $n$-block entropy per letter}. Figs. 5b show that {\it the average uncertainty per letter of an $n$-block is larger in the noise than in the  excerpts B1, B2, and B3 is}. 

\textbf{The conditional entropy}. Figs. 5c illustrate the conditional entropies, $h_n$, as a function of the word length 
$n$ for the excerpts under study.  
The noise N has significantly higher $h_n$-values.

{\it This means that the uncertainty of predicting a state one step in the future,
provided a history of the present state and the previous $n-1$ states is higher in the case of noise 
N or, in terms of predictability, the average predictability of the state following 
after a measured $n$-trajectory is higher in the bursts B1, B2, and B3. We recall that
any long-range memory decreases the conditional entropies 
and improves the changes for predictions.} 

\textbf{The entropy of the source}. One important conjecture, due essentially to Ebeling and Nicolis (1992), states that the most 
general (asymptotic) scaling of the block entropies takes the form

\begin{equation}
H(n) = e + nh + g n^{\mu_{0}}(ln n)^{\mu_{1}}
\end{equation}

where e and g are constants and $\mu_{0}$ and $\mu_{1}$ are constant exponents. 

The former equation is transformed to the simple linear relation

\begin{equation}
H(n) = e + nh 
\end{equation}

due to the observed rather linear scaling in Figs. 5a.  The associated slope $h$ could be considered as the limit entropy, 
which gives the Kolmogorov-Sinai entropy. The former takes values from zero to $\ln2$. 
Thus, one can normalize dividing the $h$-values by $\ln2$ obtaining a percentage.

Fig. 5a (upper panel) show that there is a 
clear-cut distinction of the values of the slopes $h$, that suggest a significant difference in the corresponding Kolmogorov-Sinai entropies (see Fig 5d, upper panel).

We focus on the entropies estimated by gliding (Fig. 5, lower panel).

In Fig. 5d we show the Kolmogorov- Sinai entropy estimated by:

(i) the slope of $H_n$ versus $n$ (Fig. 5a). The associateD values of Kolmogorov-Sinai entropy are shown by the solid columns in Fig. 5d. 

(ii) using the relation 

\begin{equation}
h =  \lim_{n \rightarrow \infty} h^{(n)}
\end{equation}

via the asymptotic behaviour of the Shannon $n$-block entropy per letter depicted in Fig. 5b). These Kolmogorov-Sinai-entropy values are shown in Fig. 5d (dotted columns). We clarify that the estimated values are indicative ones.

We systematically observe a drop of the entropy of the source in the bursts B1, B2 and B3. 

{\it This implies that the average amount 
of information necessary to predict the next symbol when being informed about the complete 
pre-history of the system significantly decreases in the emerged candidate precursors in respect to the noise.}  

The question which arises is whether the observer linear scaling in Figs. 5a is an algorithmic law of nature. 
This is an open problem. 

In summary, the various kind of block entropies, which quantifies dynamic aspects of a time series in a statistical manner, 
sensitively recognizes and discriminates the emerged strong EM precursors from the background noise. They suggest that
the memory (or compressibility) in the bursts B1, B2, and B3 is significantly larger in comparison to that of the noise N.  

\section{Principles of non-extensive Tsallis entropy}

In Introduction, we explained why physical systems that are characterized by long-range 
interactions, long-term memories, or multi-fractal nature, are best described by a generalized statistical mechanics' formalism 
that was proposed  by Tsallis (1998, 2009). More precisely, inspired by multifractals concepts, 
Tsallis (1988, 1998) proposed a 
generalization of the Boltzmann-Gibbs (B-G) statistical mechanics. He introduced an entropic expression characterized by 
an index $q$ which leads to a non-extensive statistics,

\begin{equation}
S_{q}=k\frac{1}{q-1}\left(1-\sum_{i=1}^{W}p_{i}^{q}\right), 
\end{equation}

where $p_{i}$ are the probabilities associated with the microscopic configurations, $W$ is their total number, 
$q$ is a real number, and $k$ is Boltzmann's constant. 

The entropic index $q$ describes the deviation of Tsallis entropy from the standard Boltzmann-Gibbs one. Indeed, 
using $p_{i}^{(q-1)}=e^{(q-1)\ln(p_{i})}\sim 1+(q-1)\ln(p_{i})$ in the limit $q\rightarrow 1$, we recover the 
usual Boltzmann-Gibbs entropy

\begin{equation}
S_{1}=-k\sum_{i=1}^{W}p_{i}\ln(p_{i}).
\end{equation}

This is the basis of the so called non-extensive statistical mechanics, which generalizes the Boltzmann-Gibbs theory. 

The entropic index $q$ characterizes the degree of non-aextensivity reflected in the following pseudo-additivity rule: 

\begin{equation}
S_{q}(A+B)=S_{q}(A)+S_{q}(B)+(1-q)S_{q}(A)S_{q}(B).
\end{equation}

For subsystems that have special probability correlations, extensivity 

\begin{equation}
S_{B-G} = S_{B-G}(A) + S_{B-G}(B)
\end{equation}

is not valid for $S_{B-G}$, but may occur for $S_{q}$ with a particular value of the index $q$. Such systems are sometimes referred to as 
non-extensive (Tsallis, 1998; 2009).

The cases $q>1$ and $q<1$, correspond to sub-additivity, or super-additivity, respectively. 
We may think of $q$ as a bias-parameter: $q<1$ privileges rare events, while $q>1$ privileges salient events (Zunino et al., 2008). 

We empasize that the parameter $q$ itself is not a measure of the complexity of the system but 
measures the degree of non-extensivity of the system. 
It is the time variations of the Tsallis entropy for a given $q$ ($S_{q}$) 
that quantify the dynamic changes of the complexity of the system. 
Lower $S_{q}$ values characterize the portions of the signal with lower complexity. 

In terms of symbolic dynamics the Tsallis entropy for the word length $n$ is (Kalimeri et al., 2008):

\begin{equation}\label{eq:11}
S_{q}(n)=k\frac{1}{q-1}\left(1-\sum_{(A_{1},A_{2},\ldots,A_{n})}[p(n)_{A_{1},A_{2},\ldots,A_{n}}]^{q}\right).
\end{equation}

\subsection{Application to the data}

Tsallis entropies are computed using the technique of lumping for binary partition (with the mean value as threshold) 
and block (word) length $n=2$. A detail calculation of the Tsalis entropies is given in Kalimeri et al., (2009).

{\it As Tsallis (1998) has already pointed out, the results depend upon the entropic index $q$ and it is expected that, for every specific use, better discrimination will achieved with appropriate 
ranges of values of $q$. The appropriate choice of this parameter remains an open problem}. We focus on this key problem. 

Recently, Solotongo-Costa and Posadas (2004) introduced a model for earthquke dynamics, which has been rooted in a nonextensive framework starting from first principles. The authors obtained 
the following analytic expression for the magnitude distribution of EQ:

\begin{equation}
\log(N(m>))=\log N+\left(\frac{2-q}{1-q}\right)\times
\log\left[1+\alpha(q-1)\times(2-q)^{(1-q)/(q-2)}\times
10^{2m}\right]
\end{equation}

where $N$ is the total number of EQs, $N(m>)$ the number of EQs
with magnitude larger than $m$, and $m\approx \log (\varepsilon)$.
This is not a trivial result, and incorporates the characteristics
of nonextensivity into the distribution of EQ by magnitude.
$\alpha$ is the constant of proportionality between the EQ energy,
$\varepsilon$, and the size of fragment, $r$. More precisely, Solotongo-Costa and Posadas assume that $\varepsilon \propto r$.

Solotongo-Costa and Posadas (2004) and Vilar et al. (2007) successfully tested the viability of this distribution function with data in various different areas. The associated nonextensive parameter 
is distributed from in the narrow area from 1.60 up to 1.71.

Importantly, we showed that the above mentioned nonextensive model also describes the sequence of pre-seismic kHz EM fluctuations (EM EQs) that detected prior to the Athens EQ (M = 5.9, September 7, 1999) (Papadimitriou et al., 2008). The associated parameter $q$ is 1.80. In part II of this contribution we will show that the recorder kHz EM fluctuations prior to the L'Aquila EQ also described by the model proposed by Solotongo-Costa and Posadas with a $q$-parameter equal to 1.82.

{\it It is very interesting to observe the similarity in the $q$-values
for all the catalogs of EQs used, as well as for 
the precursory sequences of kHz EM EQs associated with the activation of the Athens and L'Aquila faults.}
The aspects of self-affine nature of faulting and fracture are well documented. The former finding is in full agreement with these fundamentals aspects (see Part II).

Based on the aforementioned concepts, we estimate the Tsallis entropies with $q$-parameter equal to 1.8. Figure 6 shows that the Tsallis entropies in the emerged strong EM bursts drop to lower values in comparison to that of the noise. This suggests that in the noise there are many kinds of patterns, while into the bursts there are fewer patterns. 

Fig. 7b shows the Tsallis entropy into the excerpts under study. The Shannon entropy is also shown, foe comparison reason. The Shannon entropy to clearly discriminate the anomalies from the noise. However, in the case where the bursts are responses of an underlying process which is characterized by by long-range interactions, long-term memories, or multi-fractal nature, the Shannon entropy has not any physical background. 
 
\section{$T$-entropy of a string}

$T$-entropy is a novel grammar-based complexity / information measure defined for finite strings of symbols
(Ebeling et al., 2001; Tichener et al., 2005). It is a weighted count of the number of production steps required 
to construct the string from its alphabet. {\it Briefly, it is based on the intellectual economy 
one makes when rewriting a string according to some rules}.
 
An example of an actual calculation of the $T$-complexity for a finite string is given in Ebeling et al.( 2001). We briefly describe how the $T$-complexity is computed for finite strings.

The $T$-complexity of a string is defined by the use of one recursive hierarchical pattern copying (RHPC) algorithm. 
It computes the effective number of $T$-augmentation steps required to generate the string. The $T$-complexity may 
be thus computed effectively from any string and the resultant value is unique. 

The string $x(n)$ is parsed to derive constituent patterns $p_{i} \in A^{+}$ and associated copy-exponents $k_{i} \in N^{+}, i= 1,2,...,q,$ where $q \in N^{+}$ satisfying:

\begin{equation}
x = p^{k_{q}}_{q} p^{k_{q-1}}_{q-1} ... p^{k_{i}}_{i}...p^{k_{1}}_{1} \alpha_{0},
\quad \alpha_{0} \in A. 
\end{equation}

Each pattern $p_{i}$ is further constrained to satisfy:

\begin{equation}
p_{i} = p^{m_{i,i-1}}_{i-1}p^{m_{i,i-2}}_{i-2} ... p^{m_{i,j}}_{j} ... 
p^{m_{i,1}}_{1} \alpha_{i}, 
\end{equation}

\begin{equation}
\quad \alpha_{i} \in A \quad and \\
\quad 0 \leq m_{i,j} \leq k_{j}. 
\end{equation}

The $T$-complexity $C_{T}(x(n))$ is defined in terms of the copy-exponents $k_{i}$:

\begin{equation}
C_{T}(x(n)) = \sum^{q}_{i} \ln(k_{i}+1). 
\end{equation}

One may verify that $C_{T}(x(n))$ is minimal for a string comprising
a single repeating character. 

The $T$-information $I_{T}(x(n))$ of the string $x(n)$ is defined as the inverse logarithmic integral of the 
$T$-complexity divided by a scaling constant $\ln2$:

\begin{equation}
I_{T}(x(n)) = li^{-1} \left( \frac{C_{T}(x(n))}{\ln 2} \right). 
\end{equation}

In the limit $n \rightarrow \infty$ we have that $I_{T}(x(n)) \leq \ln( \# A^{n})$.  
The form of the right-hand side may be recognizable as the maximum possible $n$-block entropy of Shannon's definition. The neperian logarithm implicitly gives to the $T$-information the units of nats. $I_{T}(x(n))$ is the $T$-information of string  $x(n)$. The {\em average $T$-information rate per symbol}, referred here as the average $T$-entropy of $x(n)$ and denote by $h_{T}(x(n))$, is defined along similar lines,

\begin{equation}
h_{T} (x(n)) = \frac{I_{T}(x(n))}{n} (nats/symbol).
\end{equation}

\subsection{Application to the data}

Figure  8 shows that the average T-entropies in the emerged kHz  EM activity dramatically drop to lower values. 

{\it This indicates that a significantly lower number of production steps are required in order to construct the string 
from its alphabet into the emerged strong EM bursts. In others words, the bursts are characterized by a considerably 
lower complexity in comparison to that of the normal epoch (EM background).}

To summarize, all the used tools that are rooted on the notion of symbolic dynamics with sensitivity discriminate and recognize the kHz EM anomalies from the EM background. All the applied methods lead to the conclusion that the kHz EM bursts that emerged a few tens of hours prior to the L'Aquila EQ occurrence are characterized by a significantly lower complexity (or higher organization, higher predictability, lower uncertainty, higher compressibility) in respect to that of the EM background (noise). 

A question arises whether other tools absolutely referring to raw data and not to corresponding symbolic sequences also 
lead to the above mentioned conclusion. In the following section we analyze the data by means of approximate entropy. 

\section{Approximate entropy}

Related to time series analysis, $ApEn$ provides a measure of the degree of irregularity 
or randomness within a series of data (of length $N$). $ApEn$ was pioneered by Pincus as a measure of system complexity (Pincus, 1991). 
It was introduced as a quantification of regularity in relatively short and noisy data. It is rooted in the work of Grassberger 
and Procaccia (1983) and has been widely applied in biological systems 
(Pincus and Goldberger, 1994; Pincus and Singer, 1996 and references therein). 

The approximate entropy examines time series for similar epochs: more similar and more frequent epochs 
lead to lower values of $ApEn$. 

In a more qualitative point of view, given $N$ points, 
the $ApEn$-like statistics is approximately equal to the negative logarithm of 
the conditional probability that two sequences that are similar for $m$ points remain similar, 
that is, within a tolerance $r$, at the next point. Smaller $ApEn$-values indicate a greater 
chance that a set of data will be followed by similar data (regularity), thus, smaller values 
indicate greater regularity. Conversely, a greater value for $ApEn$ signifies a lesser chance of 
similar data being repeated (irregularity), hence, greater values convey more disorder, randomness 
and system complexity. Thus a low / high value of $ApEn$ reflects a high / low degree of regularity. 
Notably, $ApEn$ detects changes in underlying episodic behaviour not reflected in peak occurrences or 
amplitudes (Pincus and Keefe, 1992). 

The following is a description of the calculation of $ApEn$. Given any sequence of data 
points $u(i)$ from $i = 1$ to $N$, it is possible to define vector sequences $x(i)$, which consists 
of length $m$ and are made up of consecutive $u(i)$, specifically defined by the following:

\begin{equation}
x(i)=(u[i], u[i+ 1],...,u[i+m-1]).
\end{equation}

In order to estimate the frequency that vectors $x(i)$ repeat themselves throughout the data set within a tolerance $r$, the distance $d(x[i], x[j])$ is defined as the maximum difference between the scalar components $x(i)$ and $x(j)$. Explicitly, two vectors $x(i)$ and $x(j)$ are ``similar'' within the tolerance or filter $r$, namely $d(x[i],x[j]) \leq r$, if the difference between any two values for $u(i)$ and $u(j)$ within runs of length $m$ are 
less than $r$ (i.e. $|u(i + k)-u(j + k)| \leq r$ for $0 \leq k \leq m$). Subsequently, $C_{i}^{m}(r)$ is defined as the frequency of occurrence of similar runs $m$ within the tolerance $r$: 

\begin{displaymath}
C_{i}^{m}(r)=\frac{[\mbox{number of}\quad j \quad\mbox{such that}\quad d(x[i],x[j]) \leq r]}{(N-m-1)},
\end{displaymath}

where $j \leq (N-m-1)$.

Taking the natural logarithm of $C_{i}^{m}(r)$, $\Phi^{m}(r)$ is defined as the average of $ln(C_{i}^{m}(r))$:

\begin{equation}
\Phi^{m}(r)=\sum_{i} ln C_{i}^{m}(r)/(N-m-1)
\end{equation}

where $\sum_{i}$ is a sum from $i = 1$ to $(N-m-1)$. $\Phi^{m}(r)$ is a measure of the prevalence of repetitive patterns of length $m$ within the filter $r$.

Finally, approximate entropy, or $ApEn(m,r,N)$, is defined as the natural logarithm of the relative prevalence of repetitive patterns of length $m$ as compared with those of length $m+1$:

\begin{equation}
ApEn(m,r,N) = \Phi^{m}(r)- \Phi^{m+1}(r).
\end{equation}

Thus, $ApEn(m,r,N)$ measures the logarithmic frequency that similar runs 
(within the filter $r$) of length $m$ also remain similar when the length of the run is 
increased by 1. Small values of $ApEn$ indicate regularity, given that $i$ increasing run 
length $m$ by 1 does not decrease the value of $\Phi^{m}(r)$ significantly 
(i.e., regularity connotes that $\Phi^{m} [r] \approx \Phi^{m+1}[r]$). 
$ApEn(m,r,N)$ is expressed as a difference, but in essence it represents a ratio; 
note that $\Phi^ {m}[r]$ is a logarithm of the averaged $C_{i}^{m}(r)$, and the 
ratio of logarithms is equivalent to their difference. A more comprehensive description of 
$ApEn$ may be found in (Pincus, 1991; Pincus and Goldberger, 1994; Pincus and Singer,1996).

{it\ In summary, $ApEn$ is a ``regularity statistics'' that quantifies 
the unpredictability of fluctuations in a time series. The presence of repetitive patterns of 
fluctuation in a time series renders it more predictable than a time series in which such patterns are absent. 
A time series containing many repetitive patterns has a relatively small $ApEn$; a less predictable (i.e., more complex) 
process has a higher $ApEn$}.

\subsection{Application to the data}

Figure 7a shows that the approximate entropy into the three excerpts of the emerged kHz EM activity prior to the L'Aquila EQ clearly drop to 
lower values in comparison to that of the noise. 

{it\ This suggests that the candidate kHz EM precursors are governed by the 
presence of repetitive patterns which render them more predictable than noise in which such repetitive patterns are absent}. 
This result verifies the conclusions extracted by the tools of symbolic dynamics.

In oder to extract more different information probably hidden in the recorded kHz EM anomaly, in the next section we analyze the data in terms of fractal spectral analysis. 

\section{Analysis in terms of fractal spectral analysis}

It is well known that during the EQ preparation process,
the complex system of the Earth's manifests itself in linkages between space 
and time, producing characteristic fractal structures. It is expected that these fractal structures 
are mirrored in signals rooted in the EQ generation.
If a time series is a temporal fractal then a power-law of 
the form $S(f) \propto f^{-\beta}$ is obeyed, with $S(f)$ the power spectral density and $f$ the frequency. 
The spectral scaling exponent $\beta$ is a measure of the strength of time correlations. The goodness of the 
fit of a time series to the power-law is represented by the linear correlation coefficient, $r$, of this representation.

Our attention is directed to whether distinct alterations in the associated scaling parameters emerge in kHz EM fluctuations.
For this purpose, we applied the wavelet analysis technique in order to derive the coefficients of its power spectrum. 
The wavelet transform provides a representation of the signal in both the time and frequency domains. 
In contrast to the Fourier transform, which provides the description of the overall regularity of signals, 
the wavelet transform identifies the temporal evolution of various frequencies (i.e. as a time-frequency plane 
that indicates the frequency content of a signal at every time). The decomposition pattern of the time-frequency plane is 
predetermined by the choice of the basis function. In the present study, we used the continuous wavelet transform with the 
Morlet wavelet as the basis function. The results were checked for consistency using 
the Paul and DOG mother functions (Torrence and Compo, 1998).

The power spectral densities were estimated using a moving window of 256 samples and an overlap of 255 samples.  
Finally, the spectral parameters $r$ and $\beta$ were calculated for each window. 

Fig. 9 shows that in the emerged strong kHz EM bursts on April 4 2009 the coefficient $r$ 
takes values very close to 1, i.e., the fit to the power-law is excellent. This means 
that the fractal character of the underlying processes and structures is solid. This feature is 
compatible with a system close tp critical point. 

The $\beta$ exponent takes high values, i.e., between 2 and 3 into the strong EM fluctuations. This fact mirrors that:

(i) {\it The included EM events have long-range temporal correlations, namely, strong memory: 
the current value of the precursory signal is correlated not only with its most recent values 
but also with its long-term history in a scale-invariant, fractal manner. Briefly, the data indicate an 
underling mechanism of high organization that is compatible with the last stage of EQ generation.} 
 
(ii) {\it The spectrum manifests more power at lower frequencies than at high frequencies. 
The enhancement of lower frequency fluctuations physically reveals  a predominance of 
larger fracture events. This footprint is also in harmony with the final step of EQ preparation.}

(iii) {\it Two classes of signal have been widely used to model stochastic fractal time series 
(Heneghan and McDarby, 2000): fGn and fBm. 
For the case of the fGn model the scaling exponent $\beta$ lies between -1 and 1, while the fBm regime is 
indicated by $\beta$ values from 1 to 3 (Heneghan and McDarby, 2000). 
The $\beta$ exponent successfully identifies the candidate precursory activities from the 
EM noise. Indeed, the $\beta$ values in the EM background are between 1 and 2 indicating that 
the time profile of the EM time series during the quiet periods is qualitatively analogous to fGn class. 
On the contrary, the $\beta$ values in the candidate EM precursors are between 2 and 3, suggesting that the 
profile of the time series associated with the candidate precursors is qualitatively analogous to fBm class.}

We concentrate on the previous result. We emphasize that:

(i) Theoretical and laboratory experiments support the consideration that both the temporal 
and spatial activity can be described as different cuts in the same underlying fractal (Maslov et al., 1994;
 Ponomarev et al., 1997). A time series of major historical events could have temoral and spatial correlations. 

(ii)It has been pointed out that fracture surfaces can be represented by self-affine fractional Brownian 
surfaces over a wide range (Huang and Turcotte, 1988). 

{\it These two suggestions lead to the hypothesis that the fBm type profile of the precursory EM time series 
reflects the slipping of two rough and rigid Brownian profiles one over the other 
that leaded to the L'Aquila EQ nucleation.} 

In the part II of this contribution this consideration is f investigated in details. 

The $\beta$ exponent is related to the Hurst exponent, $H$, by the formula [{\it Turcotte,} 1997]:

\begin{equation} 
\beta = 2 H +1 
\end{equation}

with $0 < H < 1$ ($1 < \beta < 3$) for the fractional 
Brownian motion (fBm) model (Heneghan and McDarby, 2000). 
The exponent $H$ characterizes the persistent / anti-persistent properties of the signal. 

The range $0.5 < H < 1$ ($2 < \beta < 3$) indicates persistency, which means that if the 
amplitude of fluctuations increases in a time interval it is likely to continue increasing 
in the immediately next interval. We recall that $\beta$ values in the candidate 
EM precursors are between 2 and 3. 
The $H$ values are close to the value $0.7$ in the strong excerpts of the kHz EM activity. 
This means that the included EM fluctuations are positively correlated or persistent.   
This suggests that the underlying dynamics is governed by 
a positive feedback mechanism, which tends to lead the system out of equilibrium under 
external effects (Telesca and Lasaponara, 2006). The system acquires a self-regulating character and 
to a great extent the property of irreversibility, 
one of the important components of prediction reliability (Morgounov, 2001). Sammis and Sornette (2002) have recently presented the most important positive feedback mechanisms. 

It is expected that a positive feedback mechanism results in finite-time-singularity. The kHz EM under study seems to 
show a such behaviour by means of ``cumulative Benioff type EM energy  release'' (see Part II). Notice, a clear finite-time-singularity in terms of ``cumulative Benioff type EM energy  release''  has been reported in the case of the Athens EQ (Kapiris et al., 2004).

{\it Remark}. The $H$ exponent also reveals the ``roughness'' of the time series. 
We pay attention to the fact that the $H$ values in the strong kHz EM fluctuations are close to the value 0.7. 
Fracture surfaces were found to be self-affine over a wide range of length scales (Mandelbrot 1982). 
The Hurst exponent $H \sim 0.75$ has been 
interpreted as a universal indicator of surface roughness, weakly dependent on the nature of 
the material and on the failure mode (Lopez and Schmittbuhl, 1998; Hansen and 
Schmittbuhl, 2003; Ponson et al., 2006). {\it Thus, the universal spatial roughness of 
fracture surfaces is pretty mirrored in the roughness of the profile of strong pre-seismic kHz anomalies that have been emerged prior to the L'Aquila EQ}.

We recall that Hurst proposed the R/S method in order to identify, through the $H$ exponent, 
whether the dynamics is persistent, anti-persistent or uncorrelated. 
We wonder whether this method verifies the values of $H$-exponent estimated by the fractal spectral analysis.
We also wonder whether the hypothesis that the emerged anomaly is further supported by another method, for example, Detrended Fractal Analysis.
In the two next sections we are dealing with the aforementioned two questions.

\section {Rescaled Range Analysis: The Hurst Exponent} 

The Rescaled Range Analysis (R/S), which has been introduced by Hurst, attempts to 
find patterns that might repeat in the future. Briefly, there are two main variables used 
in this method, the range of the data (as measured by the highest and lowest values in the time period), 
and the standard deviation of the data. 

Hurst, in his analysis, first transformed the natural records in time $X(N)=x(1),x(2),...,x(N)$, into a new variable $y(n,N)$, the so-called accumulated departure of the natural record in time in a given year $n (n = 1,2,...N)$, from the average, $<x>(n)$, over a period of $N$ years. The transformation follows the formula 

\begin{equation}
y(n,N) = \sum_{i=1}^{n} (x(i) - \langle x \rangle )
\end{equation}

Then, he introduced the rescaled range

\begin{equation}
R/S = \frac{R(N)}{S(N)}
\end{equation}

in which the range $R(N)$ is defined as a distance between the minimum and maximum value of $y$ by 

\begin{equation}
R(N) = y_{max} - y_{min}
\end{equation}

and the standard deviation $S(N)$ by

\begin{equation}
S(N) = \sqrt{\frac{1}{N} \sum_{i=1}^{N} [y(i) - \langle x \rangle] ^2}
\end{equation}

R/S is expected to show a power-law dependence on the box size $n$:

\begin{equation}
R(n)/S(n) \sim nH , 
\end{equation}

where $H$ is the Hurst exponent.

\subsection{Application to the data}

Fig. 7 shows that the R/S technique applied directly on the raw data may be of use in distinguishing 
``candidate pathological'' from ``healthy'' data sets in terms of $H$ exponent. The ``healthy'' data 
(EM background) are characterized by antipersistency. In contrast, the ``candidate pathological'' data sets 
are characterized by strong persistency. 

We empasize, that the $H$ exponents derived from the relation $\beta = 2H+1$, which is valid for 
the fBm-model, follow quite nicely those estimated by the R/S analysis. 
This consistency supports the hypothesis that the candidate EM precursors follow the fBm-model.

 \section {Detrended Fluctuation Analysis} 

Often experimental data are affected by non-stationarities, and strong trends in the 
data can lead to a false detection of long-range correlations if the results are not 
carefully interpreted. The DFA, proposed by Peng et al. (1993, 1994, 1995) and based on random walk 
theory, is a well-established method for determining the scaling behaviour of noisy data in the presence 
of trends without knowing their origin and shape. 

We briefly introduce the DFA method, which involves the following six steps:
 
(i) We consider a time series of $i = 1, ..., N$ of length $N$. In most applications, the index $i$ will correspond to the time of measurements. We are interested in the correlation of the values $x_i$ and $x_{i+k}$ for different time lags, i.e. correlations over different time scales $k$. In the first step, we determine the integrated profile 

\begin{equation} \label{eq1}
y(k) = \sum_{i=1}^{k} (x(i) - \langle x \rangle ),   i=1,..,N
\end{equation}
where $\langle ... \rangle$ denotes the mean. 

(ii) The integrated signal $y(k)$ is divided into non-overlapping boxes of equal length $n$.

(iii) In each box of length $n$, we fit $y(k)$, using a polynomial function of  order $l$, which represents the trend in that box. We usually use a linear fit. The $y$ coordinate of the fit line in each box is denoted by $y_n(k)$.

(iv)  The integrated signal $y(k)$ is detrended by subtracting the local trend $y_n(k)$. Then we define the detrended time series for boxes duration $n$, denoted by $y_n(k) = y(k) - y_n(k)$.

(v) For a given box size $n$, the root-mean-square (rms) fluctuations for this integrated and detrended signal is calculated: 

\begin{equation} 
F(n) = \sqrt{\frac{1}{N} \sum_{k=1}^{N} \{y(k) - yn(k)\} ^2}
\end{equation} 

(vi) The aforementioned computation is repeated for a broad range of scales box sizes ($n$) to provide a relationship between $F(n)$ and the box size $n$.
 
A power-law relation between the average root-mean square fluctuation $F(n)$ and the box size $n$ indicates the presence of scaling:

\begin{equation} \label{eq4}
F(n)\sim n^\alpha
\end{equation} 

The scaling exponent $\alpha$ quantifies the strength of the long-range power-law correlations in the time series. 

\subsection{Application to the data}

We fit the experimental time series by the function $F(n)\sim n^\alpha$. 
In a $logF(n)-logn$ representation the former function is a line with slope $\alpha$. 
We note that the scaling exponent $\alpha$ is not always constant (independent of scale) 
and crossovers often exist, i.e., the value of $\alpha$ differs for long and short time scales. 
In order to examine the probable existence of crossover behaviour, the short-term 
scaling exponent $\alpha_1$ and the long-term scaling 
exponent $\alpha_2$ were estimated into the noise N and the exsepts B1, B2, and B3.

Following Peng et al., (1995) in Fig. 10 we show the scatter plot of scaling exponents $\alpha_1$ vs $\alpha_2$. 
The behaviour of these two exponents clearly separates the EM noise from candidate EM precursors. 
The three bursts are characterized by much larger $\alpha_1$ and $\alpha_2$ values. 
More precisely, in the noise the two exponents have values close to 1 indicating an underlying $1/f$-type noise. 
On the contrary, the three bursts show exponents higher to 1 and close to 1.5, {\it pretty close to that of a fBm ($\alpha_2 \sim 1.5$)} 
(Peng et al., 1995). 

This finding also mirrors that kHz EM fluctuations 
are governed by strong long-range power-law correlations 
inductive of an underlying positive feedback mechanism, which has a propensity to lead the system out of equilibrium under 
external effects. 

The analysis also shows that the candidate EM precursors do not exhibit a clear crossover in scaling behaviour, both the $\alpha_s$ and $\alpha_l$ exponents have values pretty close to that (1.5) of a fBm. 
Finally, the application of the DFA to data verifies that the strong kHz activities follows the fBm-model.

\section{View of candidate precursory patterns in terms of theory of complexity}

Earthquake occurrence is a complex phenomenon in space and time. 
The field of study of complex systems holds that the dynamics of complex systems is founded
on universal principles that may used to describe disparate problems ranging from
particle physics to the economics of societies (Stanley, 1999, 2000; Stanley et al., 2000; Vicsek, 2001, 2002). 

The study of complex system in a unified framework has become recognized in recent years 
as new scientific discipline, the ultimate of interdisciplinary fields. Characteristically, 
de Arcangelis et al. (2006) presented evidence for universality in solar flare and EQ occurrence. 
Picoli et. al.  (2007) reported similarities between the dynamics of geomagnetic signal and heartbeat intervals. 
Kossobokov and Keilis-Borok (2000) have explored similarities of multiple fracturing on a neutron star and on the Earth, 
including power-law energy distributions, clustering, and the symptoms of transition to a major rupture. 
Sornette and Helmstetter (2002) have presented occurrence of finite-time singularities in 
epidemic models of rupture, EQ, and starquakes. 
Kapiris et al., (2005) and Eftaxias et al., ( 2006) reported similarities 
in precursory features in seismic shocks and epileotic seizures. Osario et al. (2007) have suggested 
that that epileptic seizure could be considered as quakes of brain. Fukuda et al., (2003) reported  
similarities between communication dynamics in the Internet and the automatic nervous system.

Breaking down the barriers between physics, chemistry and biology and the so-called soft sciences of psychology, sociology economics, and anthropology, this approach explores the universal physical and mathematical principles that govern the emergence of complex systems from simple components (Bar-Yan, 1997; Sornette, 2002; Rundle et al., 1995). 

{\it One of the issues that we will need to address is whether the crucial pathological symptoms of low complexity and persistency included in the candidate kHz EM precursor also characterize other, different, however, in their nature, catastrophic events}.

We investigate the probable presence of the above mentioned pathological symptoms in epileptic seizures, 
magnetic storms, solar flares.

\subsection{Similarities between the dynamics of magnetic storm and EM precursors} 

Intense magnetic storms are undoubtedly among the most important phenomena in space physics 
interlinking the solar wind, magnetosphere ionosphere, atmosphere and occasionally the Earth's crust 
(Daglis, 2001; Daglis et al., 2003). The Dst index is a geomagnetic index which monitors the world-wide magnetic storm level. It is expressed in nanoteslas and is based on the average value of the horizontal component of the Earth's magnetic field measured hourly at four near-equatorial geomagnetic observatories. 

In recent works Balasis et al. (2006, 2008, 2009a, 2009b) studied Dst data (2001) which include intense magnetic storms, as well as a number of smaller events.  Therein, we have applied the majority of the techniques 
that have been used in the present work. The results show that all the extracted crucial features from the kHz EM activity under study (e.g., long-range correlations, persistency, appearance of fluctuations at all scales with a simultaneous predominance of large events)  are also hidden in intense magnetic storms. 
We have suggested that the development of both intense magnetic storm or kHz EM can be studied in a unified framework, i.e., within the school of the "Intermittent Criticality". This school has a more general character 
than the classical self-organized criticality, implying the predictability 
of the impending catastrophic event.

In Part II of this contribution more quantitative evidence of universal behaviour between the kHz EM precursors under study and intense magnetic storms is presented. We emphasize that based on previously detected kHz EM pre-seismic anomalies, we have already shown that the kHz EM precursors and magnetic storms share common scale-invariant natures(Papadimitriou et al., 2008; Balasis et al., 2009a). The relevant analysis is based on  
the nonextensive model of EQ dynamics presented by (Solotongo-Costa and Posadas, 2004). 

\subsection{Similarities between the epileptic seizures and EM precursors} 

Theoretical studies suggest that the final EQ and neural-seizure dynamics should have many similar features and could be analyzed within similar mathematical frameworks (Hopfield et al., 1994; Rudle et al., 1995; Herz and Hopfield, 1995). Recently, we studied the temporal evolution of the fractal spectral characteristics in: (i) electroencephalograph (EEG) recordings on rat experiments included epileptic shocks, and (ii) pre-seismic kHz EM time series detected prior to the Athens EQ.  
We showed that similar distinctive symptoms (including all the observed in 
the present study precursory symptoms) appear in epileptic seizures and EM precursors(Kapiris et al., 2005; Eftaxias et al., 2006). Importantly, we propose that these two observations also find a unifying explanation within the school of the "Intermittent Criticality".

We notice that recently, Osorio et al., (2007)  have tested the universal 
hypothesis on the Earth's crust and the epileptic brain suggesting 
that epileptic seizures could be considered as quakes of the brain.

\subsection{Similarities between solar flares and EM precursors} 

In a recent work Koulouras et al. (2009) investigated MHz EM radiations rooted in solar flares. A comparative study show that these emissions include all the extracted precursory features from the kHz EM emission under study via the fractal spectral analysis. Importantly, the solar activity follows the ``persistent fBm model'', while persistent behaviour is not found in quiet Sun observations. Schwarz et al. (1998) verify this result: they showed that the time profiles of solar mm-wave bursts are qualitatively analogous to fBm model showing persistent behaviour. Notice,  
de Arcangelis et al. (2006) presented evidence for universality in solar flare and EQ occurrence, while Papadimitriou et al. (2008) reported indications for universality in kHz pre-seismic EM activities and EQ.

In a forthcoming paper we successfully test the universal 
hypothesis on the kHz EM anomalies detected prior to the L'Aquila EQ and solar flares. The relevant analysis is also based on  
the nonextensive model of EQ dynamics presented by (Solotongo-Costa and Posadas, 2004). 

{\it In summary, the kHz EM precursors under study, epileptic seizures, solar flares, and magnetic storms include  "universal" symptoms in their internal structural patterns. These symptoms clearly recognize these catastrophic events from the corresponding normal state}. 

\section{Conclusions}

A sequence of ULF, kHz and MHz EM anomalies recorded at Zante station, from March 26 2009 up to April 4 2009, namely prior to the L'Aquila EQ that occurred on April 6 2009. Our effort is focused on the question whether the recorded anomalies are seismogenic or not. Notice, we investigate not only whether or not a single kHz, MHz, or ULF EM anomaly is pre-seismic in itself, but mainly whether a combination of such anomalies can be characterized as pre-seismic. Our approach is based on the following open key questions: How we can recognize an EM observation as a pre-seismic one. How we can link an individual EM precursor with an equivalent distinctive stage of EQ preparation. How we can identify precursory symptoms in EM observations that witness that the occurrence of the prepared EQ is unavoidable. 

In this Part I of our contribution, we restrict ourselves in the study of the recorded sequence of kHz EM anomalies. The pre-seismic kHz anomalies have a crucial role in the frame of our work (e.g Kapiris et al., 2004, Contoyiannis et al., 2005;  Eftaxias et al., 2007; Papadimitriou et al., 2008).  We attempt to approach the remaining question in Part II. 

Based on a multidisciplinary investigation, we argue that the emerged sequence of kHz anomalies clearly are recognized and discriminated by the EM background in region of the station.  In order to develop a quantitative identification of a kHz EM anomaly, concept of entropy and tools of information theory have been used in order to identify statistical patterns: a significant change in the statistical pattern represents a deviation from the normal behaviour revealing the presence of an anomaly. In principle one cannot find an optimum tool for anomaly detection. A combination of various tools seems to be a promising way to get a more precise recognition of a recorded anomaly as a pre-seismic one. We analyzed the associated kHz EM time series in terms of Shannon $n$-block entropy, differential or conditional entropy, entropy of the source or limit entropy, Kolmogorov-Sinai entropy, $T$-entropy, and nonextensive Tsallis entropy. For comparison reasons, we applied 
a more tool to the data, namely, approximate entropy which refers directly to the raw data and not to a transformed symbolic sequence. We conclude that all the applied methods sensitively recognize the launched candidate kHz EM from the normal state (EM background):
{\it they are characterized by a considerably lower complexity (or higher organization, lower uncertainty, higher predictability, higher compressibility) in comparison to that of EM background (noise)}.
The spectral fractal technique further discriminates the anomalies from the noise. {\it The candidate precursors follow the persistent fBm-model, while the noise tracks the 1/f-noise model}. We verify the existence of persistency into the anomalies by the R/S analysis. The suggestion that the candidate precursors follow the fBm-model is verified by means of detrended fractal analysis. 

The fact that the launch of anomalies from the normal state is combined by (i) a simultaneous appearance of a significantly higher organization, and (ii) persistency, indicates that the process, where the anomalies are rooted, is governed by a positive feedback mechanism. This mechanism is inductive of an out of equilibrium process, i.e., catastrophic event. 

In this field of research, the reproducibility of results is desirable. Importantly, the included catastrophic symptoms in the candidate kHz EM precursors under study are also included in rather well established kHz EM precursors associated with significant EQ that recently occurred in Greece (e.g Kapiris et al., 2004, Contoyiannis et al., 2005;  Eftaxias et al., 2007; 2008; Kalimeri et al., 2008) .    

The field of study of complex systems holds that their dynamics is rooted in universal principles that may used to describe disparate problems ranging from particle physics to the economics of societies. Evidence showed that the low complexity and persistency common footprints in other catastrophic events, namely, neural-seizures, magnetic storms, and solar flares. Part II of this communication further supports the hypothesis that kHz EM precursors, solar flares, magnetic storms and epileptic seizures could be investigated in a unified framework. 

We wonder whether the clear discrimination of the emerged prior to the L'Aquila EQ kHz EM anomalies from 
the normal state, even based on a combination of (i) a rather strong statistical analysis,  and (ii) striking similarities with 
other complex catastrophic events, reliably leads to the conclusion that these anomalies are rooted in the 
preparation of the L'Aquila EQ. 

{\it Our view is that such an analysis by itself cannot establish an 
anomaly as precursor and even more identify a corresponding characteristic stage in the EQ preparation process. It probably offers sufficient but not necessary criteria in order to recognize an anomaly as pre-seismic 
one}. 
  
Much remains to be done to tackle systematically precursors. 
It remains to be established whether different approaches could provide additional information that allows one to accept the seismogenic origin of the recorded kHz EM anomalies and correspond these to a distinguishing stage of EQ generation. 

Fracture process is characterized by fundamental universally valid scaling relationships. In particular, the aspects of self-affine nature of faulting and fracture have been well-established. Our point of view is that such fundamental structural patterns of fracture and faulting process should be reflected on a real fracto-electromagnetic activity. Thus, their potential existence in the candidate precursor would be a strong footprint of its seismogenic origin. In Part II of this communications we show the existence of such structural precursory patterns in the candidate kHz.
In part II, based on the strategy described in Introduction, we complete our study. More precisely, an important pursuit is to make a quantitative comparison between pre-seismic EM time series and fracture surfaces. We support the seismogenic origin of the detected MHz EM anomalies. We try to correlate the kHz and MHz EM anomalies with distinctive last stages of the L'Aquila preparation process. It is a risky practice to extend findings rooted in laboratory experiments to geophysical scale. However, one can not ignore a comparison between failure precursors at laboratory and geophysical scale. Finally, the reproducibility of results is desirable in this field of research. We refer to this requirement. Finally, we discuss the conditions that favourable the detection of kHz-MHz EM emissions rooted in a fracture process in the focal area. We explore this requirement in terms of fractal electrodynamics, which combines fractal geometry with Maxwell's equations. It is a difficult task to separate two events, such as a candidate precursor and the ensuing EQ, separated in time.

\newpage

\begin{figure}[t]
\vspace*{2mm}
\begin{center}
\includegraphics[width=8.3cm]{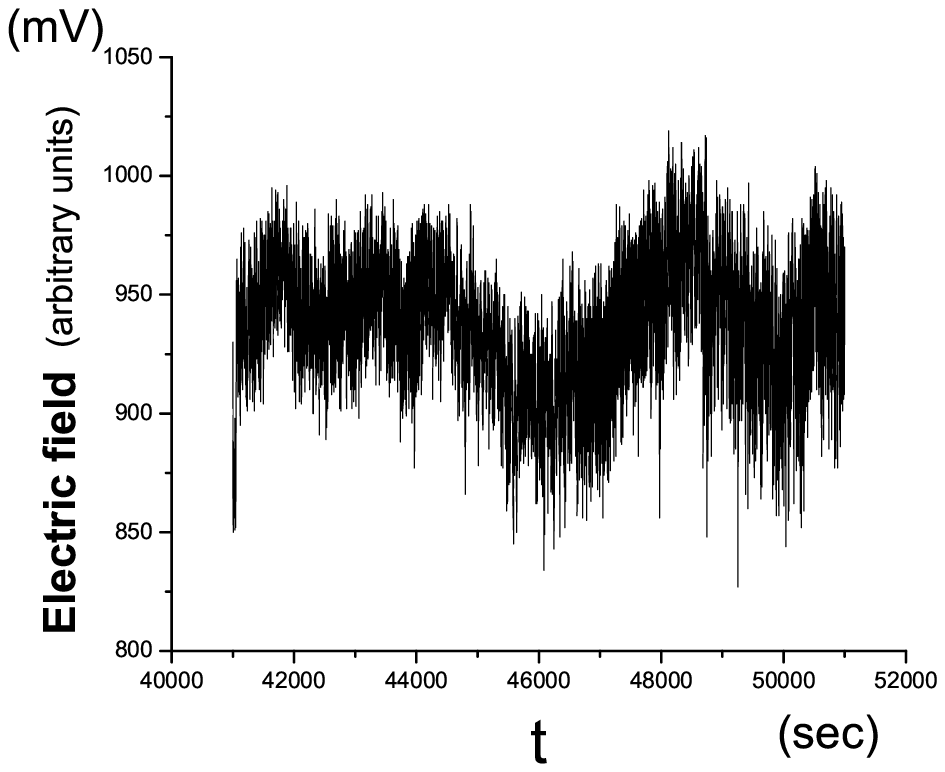}\\
\includegraphics[width=8.3cm]{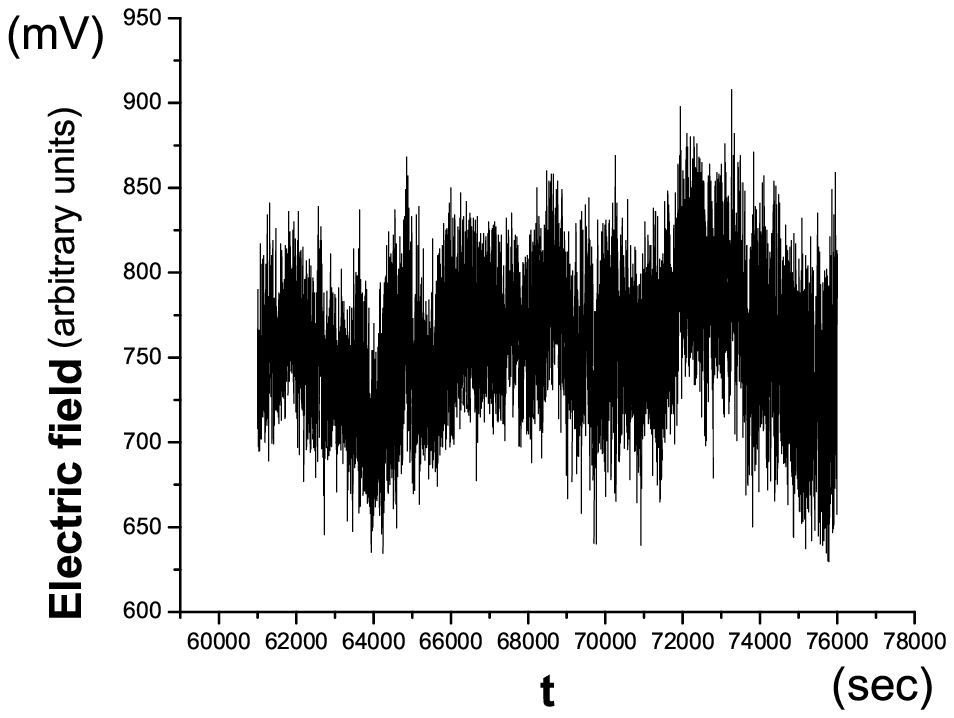}
\end{center}
\caption{The critical excerpts of the 41 MHz magnetic field strength time series on March 26 2009 (upper panel) 
and April 2 2009 (lower panel), respectively. The EM fluctuations included in each time interval behave as a 
continuous (second order) phase transition in striking similarity with the behaviour of corresponding MHz EM 
precursors detected prior to other significant EQ that recently occurred in Greece (Contoyiannis et al., 
2005). Their critical behaviour has been investigated by means of a recently introduced method of critical 
fluctuations (Contoyiannis and Diakonos, 2000; Contoyiannis, Diakonos, and Malakis, 2002).}
\end{figure}

\newpage

\begin{figure}[t]
\vspace*{2mm}
\begin{center}
\includegraphics[width=8.3cm]{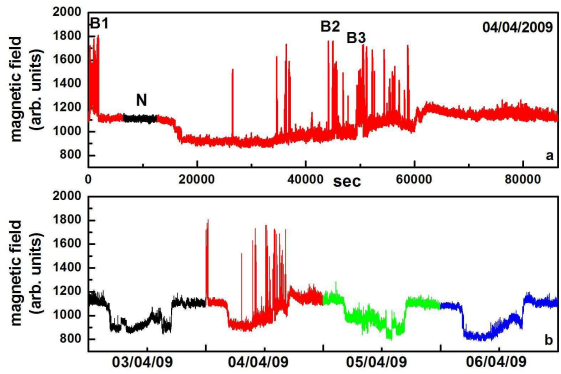}
\end{center}
\caption{(a) We observe the presence of a sequence of strong EM impulsive bursts at 10 kHz on April 4 2009. (b) 
These anomalies are launched over a quiescence period concerning the detection of EM disturbances at the kHz 
frequency band. A segment from the EM background (N) and three excerpts of the emerged strong kHz EM activity (B1, B2, B3) have been marked in the time series of April 4 2009.}
\end{figure}

\newpage

\begin{figure}[t]
\vspace*{2mm}
\begin{center}
\includegraphics[width=8.3cm]{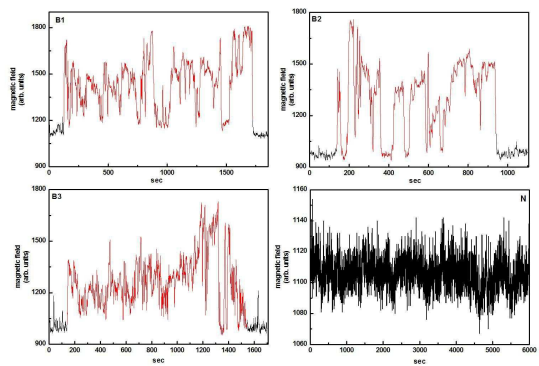}
\end{center}
\caption{Magnified images of the excerpts N, B1, B2, and B3 that are shown in Fig. 2.}
\end{figure}

\newpage

\begin{figure}[t]
\vspace*{2mm}
\begin{center}
\includegraphics[width=8.3cm]{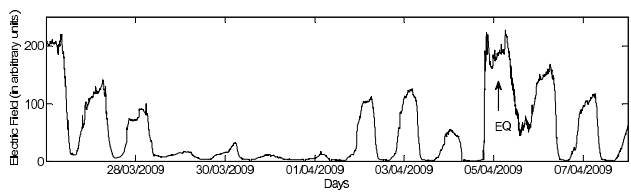}
\end{center}
\caption{The time series of the ULF electric field strength, as it was recorded by the STWA sensors.}
\end{figure}

\newpage

\begin{figure}[t]
\vspace*{2mm}
\begin{center}
\includegraphics[width=8.3cm]{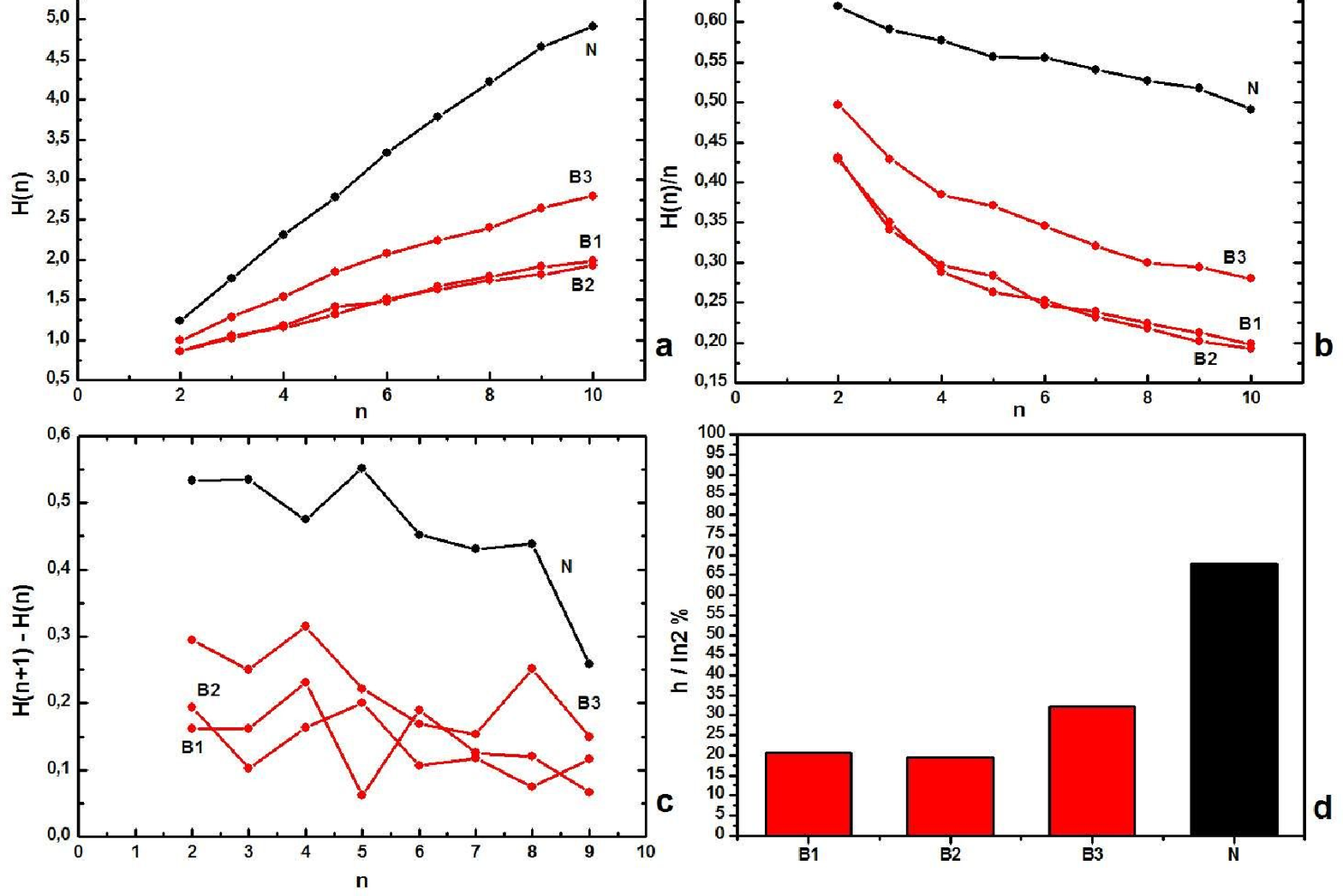}\\
\vspace*{-40mm}
\includegraphics[width=8.3cm]{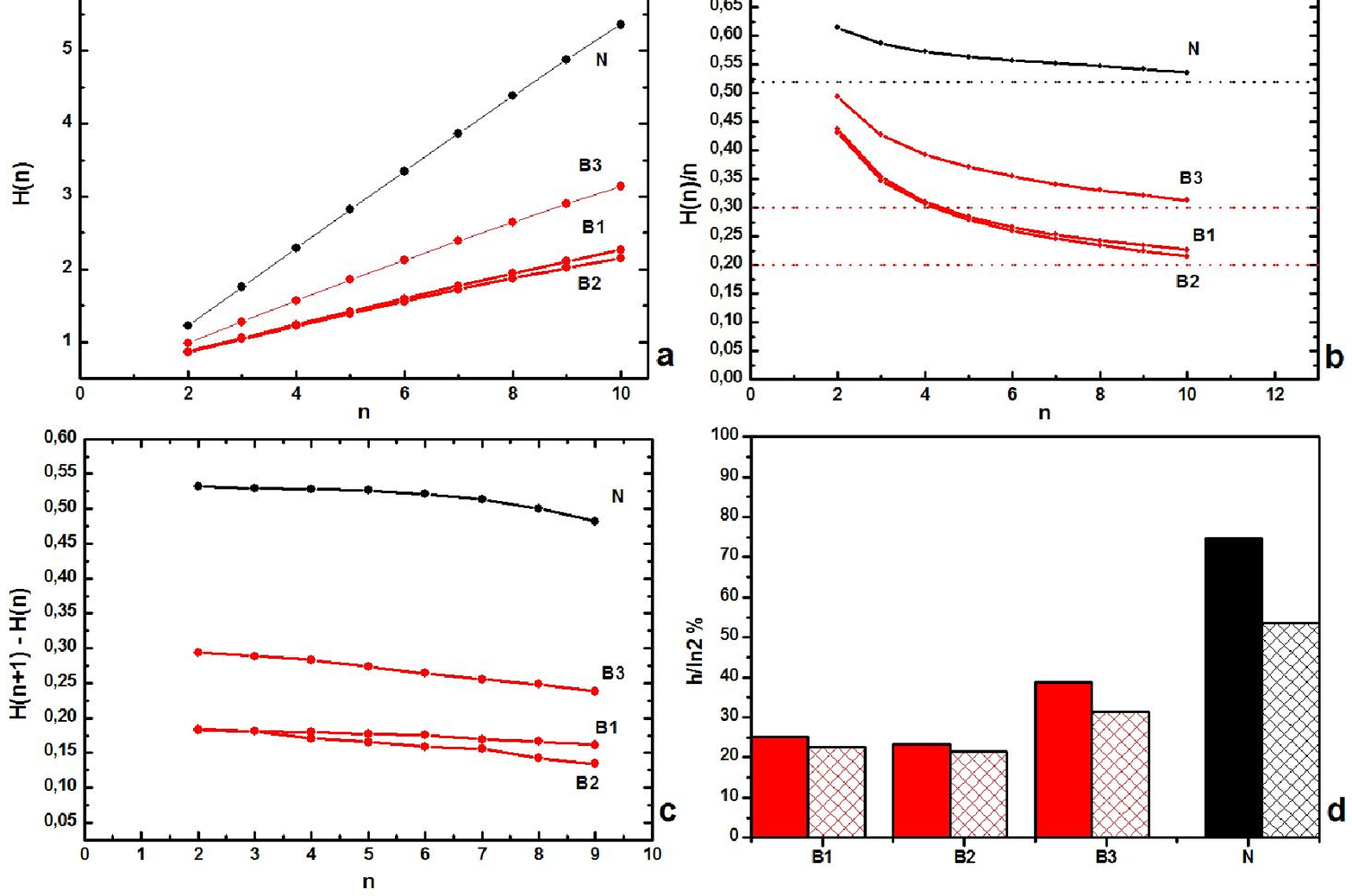}
\end{center}
\vspace*{-20mm}
\caption{The Shannon $n$-block entropy $H(n)$, conditional entropy $H(n+1) - H(n)$, Shannon $n$-block entropy per 
letter $H(n)/n$ and Kolmogorov-Sinai entropy ($h/ln2$) as a function of word length $n$, into the background noise (N) and 
the three candidate precursory EM bursts B1, B2, B3 for lumping (upper panel) and gliding (lower panel). All 
these symbolic entropies for either reading techniques show a significant drop of complexity into the EM bursts 
with respect to the noise.}
\end{figure}

\newpage

\begin{figure}[t]
\vspace*{2mm}
\begin{center}
\includegraphics[width=8.3cm]{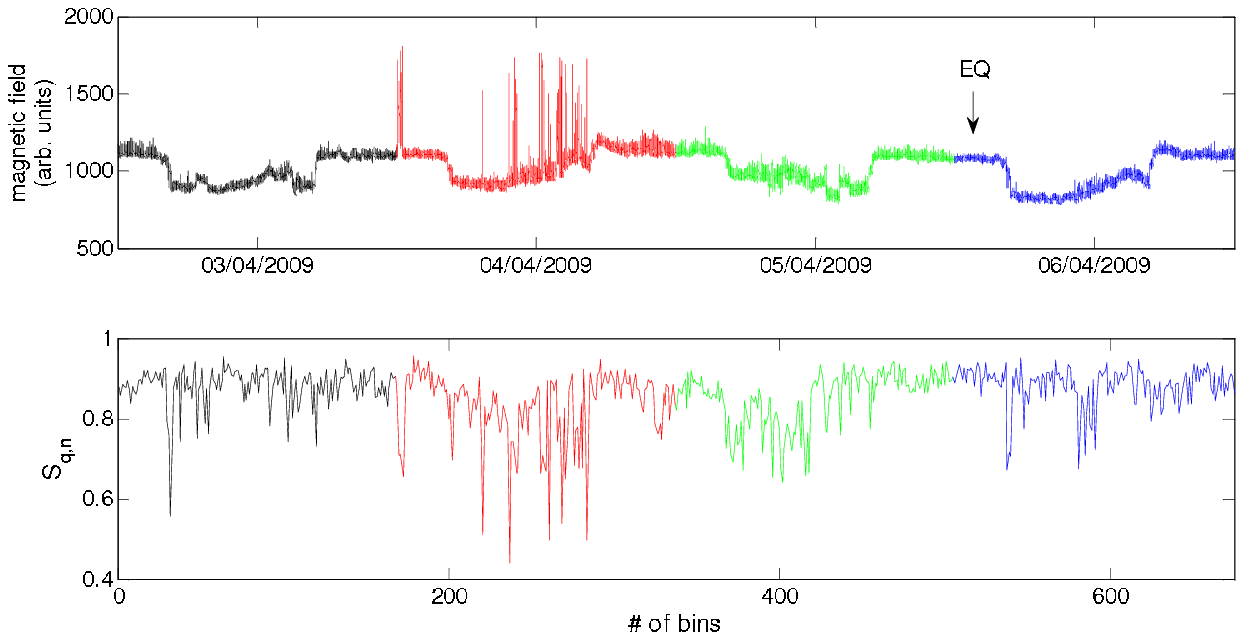}\\
\includegraphics[width=8.3cm]{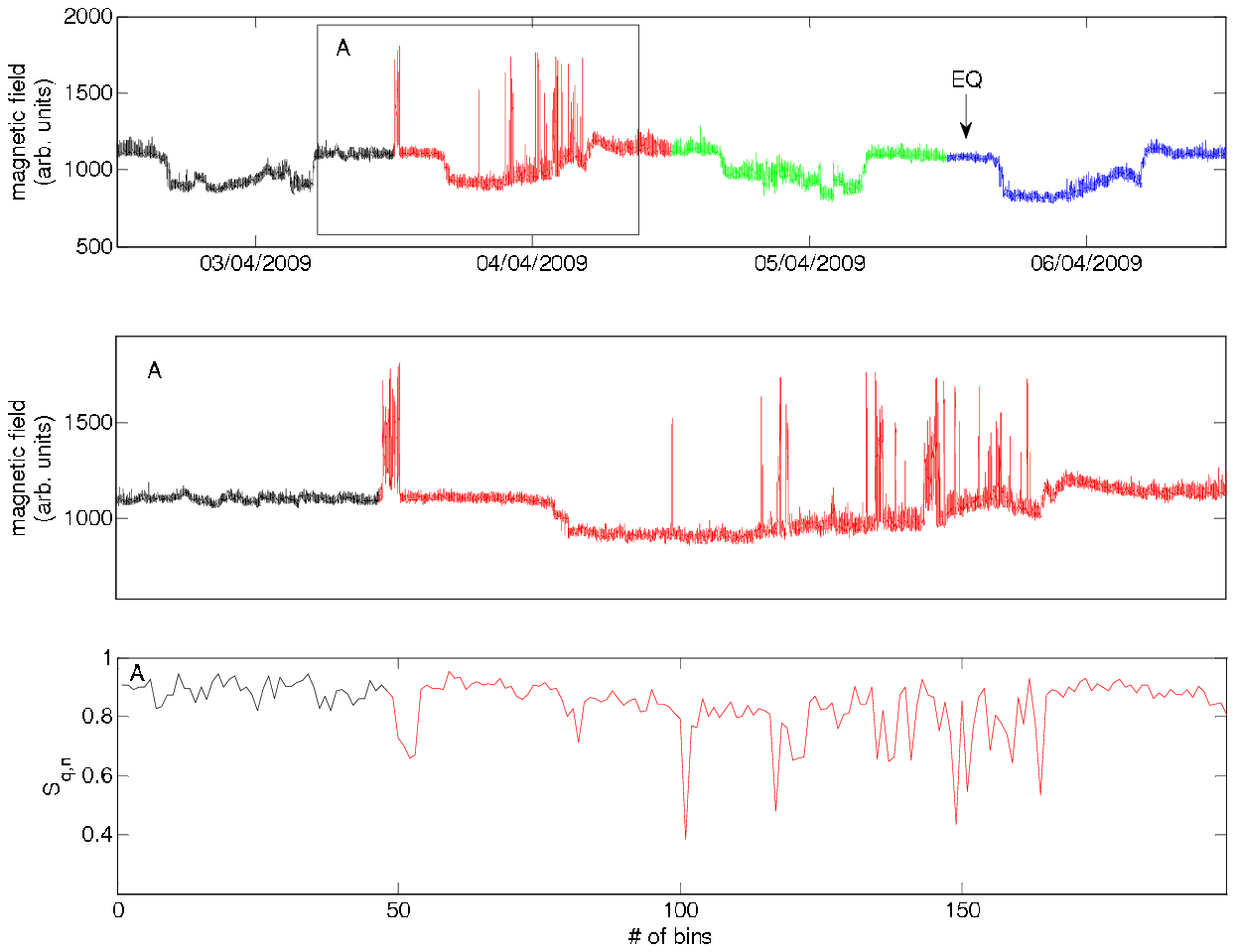}\\
\includegraphics[width=8.3cm]{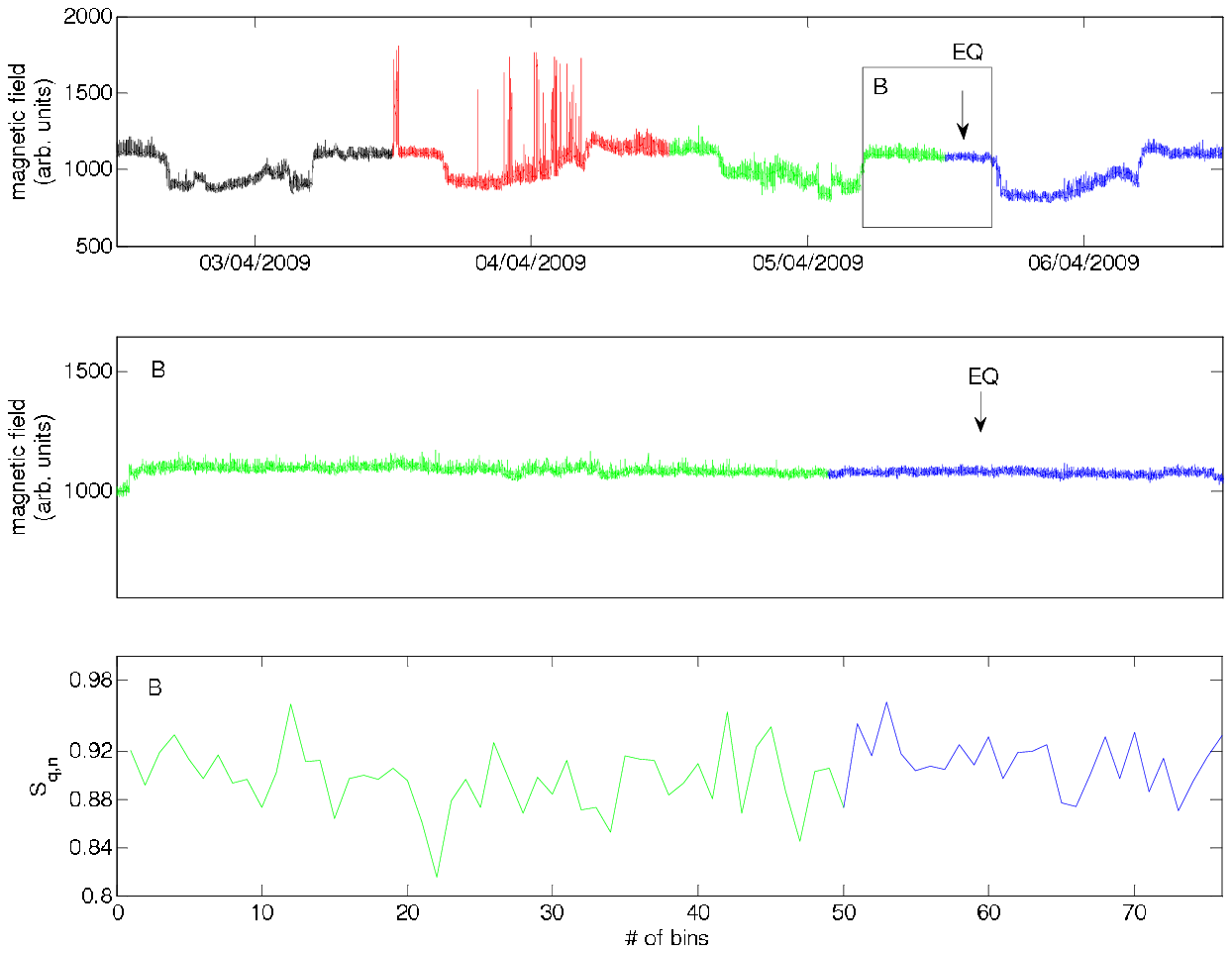}
\end{center}
\caption{The normalized Tsallis entropy has significantly lower values in the candidate EM precursors B1, B2, and B3 in comparison to that of the noise N. We recall that broad symbol-sequence frequency distributions produce high entropy values, indicating a low degree of organization. On the contrary, when certain symbol sequences exhibit high frequencies, lower entropy values are produced, indicating a high degree of organization. We conclude that the bursts B1, B2, and B3 are characterized by higher organization in respect to that of the noise N.}
\end{figure}

\newpage

\begin{figure}[t]
\vspace*{2mm}
\begin{center}
\includegraphics[width=8.3cm]{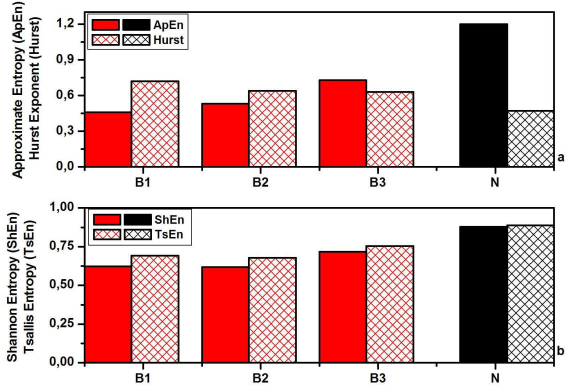}
\end{center}
\caption{For comparison reasons, we present the values of $ApEn$, Hurst exponent estimated by R/S analysis, Tsallis entropy and Shannon 
entropy for B1, B2, B3 and N. The values of $ApEn$ significantly decrease in the three EM bursts B1, B2, B3 with 
respect to the noise (N). Smaller $ApEn$-values indicate a greater chance that a set of data will be followed by 
similar data, thus, the smaller values in the EM bursts indicate greater regularity. The Hurst-exponent estimated 
by R/S analysis also clearly separates the EM noise N from candidate EM precursors B1, B2, and B3. The candidate 
EM precursors sre characterized by persistency. On the contrary, the noise N follows the fGn model. Taken together 
the higher organization indicated by the $ApEn$ and persistency indicating by $H$ exponent we conclude that the 
EM bursts are associated with a positive feedback mechanism which leads the system out of equilibrium. Shannon and 
Tsallis entropies are also indicative for higher organization in the EM bursts.}
\end{figure}

\newpage

\begin{figure}[t]
\vspace*{2mm}
\begin{center}
\includegraphics[width=8.3cm]{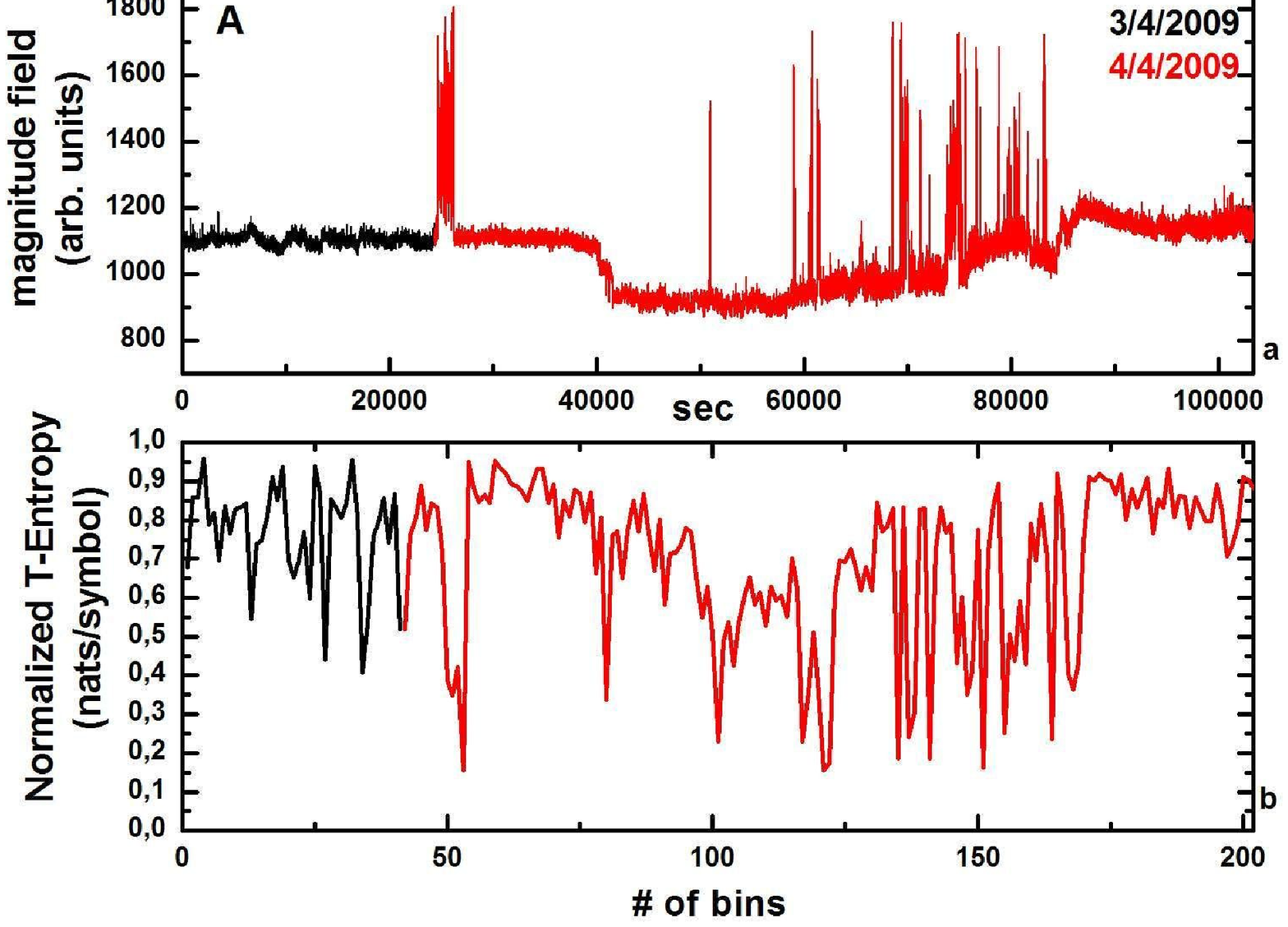}\\
\vspace*{-40mm}
\includegraphics[width=8.3cm]{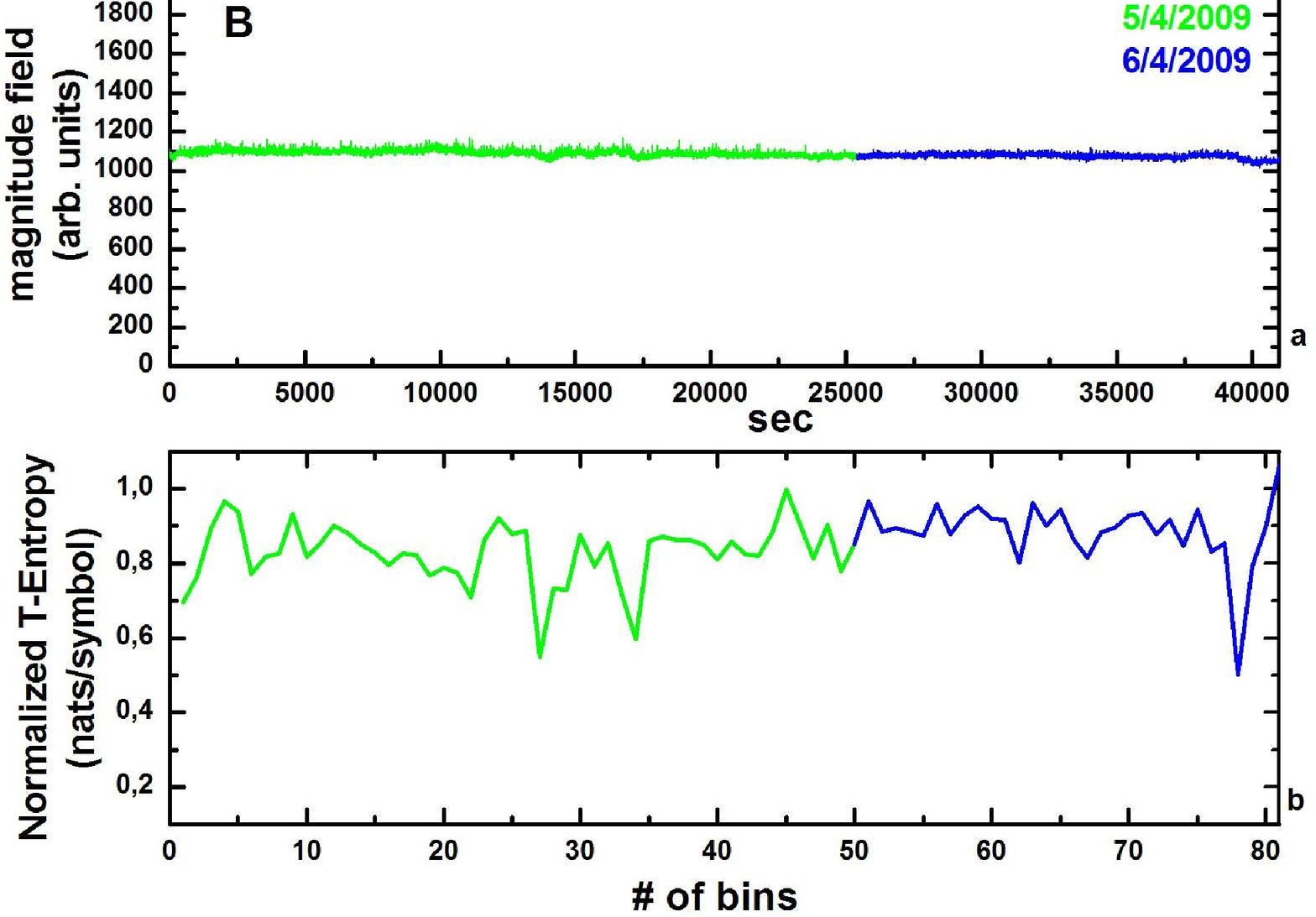}
\end{center}
\vspace*{-20mm}
\caption{Values of normalized $T$-entropy for time intervals A and B (upper and lower panel, respectively). 
Time intervals A and B are defined in Figure 6. In the case of burst we observe that less production steps are 
required in order to construct the string from its alphabet.}
\end{figure}

\newpage

\begin{figure}[t]
\vspace*{2mm}
\begin{center}
\includegraphics[width=8.3cm]{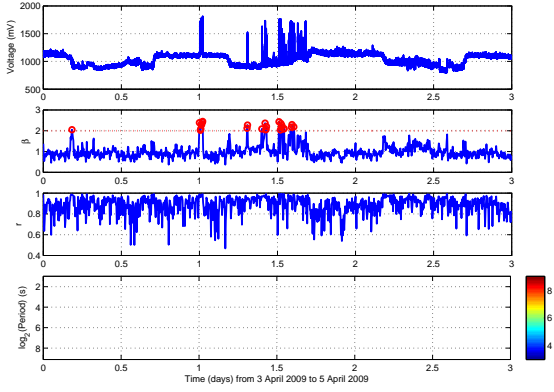}
\end{center}
\caption{From top to bottom are shown the 10  kHz time series, spectral exponents $\beta$, linear correlation coefficients $r$, and the wavelet power spectrum from 12:00:00 3 April 2009 to 12:00:00 5 April 2009. The red dashed line in the $\beta$ plot marks the transition between anti-persistent and persistent behavior. We observe that in the emerged strong kHz EM bursts the coefficient $r$ takes values very close to 1, i.e., the fit to the power-law is excellent. This means that the fractal character of the underlying processes 
and structures is solid. The $\beta$ exponent takes high values, i.e., between 2 and 3, following the fBm model. 
This fact mirrors that: (i) the included EM events have strong memory; (ii) the spectrum manifests more power at 
lower frequencies than at high frequencies, in other words a predominance of large fracture events; and (iii) 
follows the fBm model. We emphasize that the strong EM fluctuations are governed by persistency. This is verified 
by the R/S analysis (Figure 7). On the contrary, the background behaves as $1/f$-type noise.}
\end{figure}

\newpage

\begin{figure}[t]
\vspace*{2mm}
\begin{center}
\includegraphics[width=8.3cm]{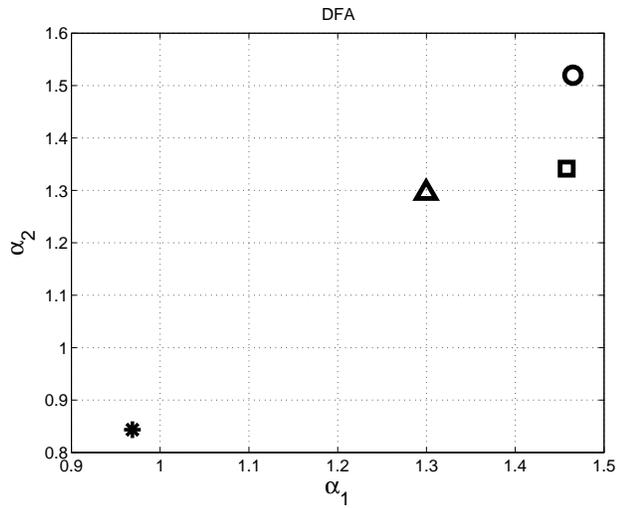}
\end{center}
\caption{The scatter plot of scaling exponents $\alpha_1$ vs $\alpha_2$ clearly separates the EM noise N (denoted 
by a star) from candidate EM precursors B1 (square), B2 (triangle), and B3 (circle). In the noise the two 
exponents have values close to 1 indicating an underlying $1/f$-type noise. On the contrary, the three bursts 
show exponents higher to 1 and close to 1.5, pretty close to that of a fBm ($\alpha_2 \sim 1.5$).}
\end{figure}

\end{document}